\documentclass[aip, jcp, reprint]{revtex4-1}  
\pdfoutput=1  

\usepackage[T1]{fontenc}
\usepackage[font=sf]{floatrow}

\makeatletter	
\renewcommand\@make@capt@title[2]{%
 \@ifx@empty\float@link{\@firstofone}{\expandafter\href\expandafter{\float@link}}%
  {\sf\textbf{#1}}\sf\@caption@fignum@sep#2\quad
}%
\makeatother

\usepackage{amsmath}
\usepackage{amssymb}
\usepackage{graphicx}
\usepackage{caption}
\usepackage{dcolumn}
\usepackage{boxedminipage}
\usepackage{verbatim}
\usepackage{booktabs}
\usepackage{subfigure}
\usepackage{listings}

\usepackage[colorlinks=true,citecolor=blue,linkcolor=blue]{hyperref}

\newcommand{\var}[1]{{\mathrm var}{(#1)}}

\begin{document}


\title{Towards Automated Benchmarking of Atomistic Forcefields:\\
Neat Liquid Densities and Static Dielectric Constants from the ThermoML Data Archive}

\author{Kyle A. Beauchamp$^+$}
\thanks{Corresponding author}
\email{kyle.beauchamp@choderalab.org}
\affiliation{Computational Biology Program, Sloan Kettering Institute, Memorial Sloan Kettering Cancer Center, New York, NY}

\author{Julie M. Behr$^+$}
\affiliation{Tri-Institutional Program in Computational Biology and Medicine, Weill Cornell Medical College, New York, NY}

\author{Ari\"{e}n S. Rustenburg}
\affiliation{Graduate Program in Physiology, Biophysics, and Systems Biology, Weill Cornell Medical College, New York, NY}
 
 \author{Christopher I. Bayly}
 \affiliation{OpenEye Scientific Software Inc., Santa Fe, NM}

 \author{Kenneth Kroenlein}
 \affiliation{Thermodynamics Research Center, NIST, Boulder, CO}
 
 \author{John D. Chodera}
 \thanks{Corresponding author}
 \email{john.chodera@choderalab.org}
 \affiliation{Computational Biology Program, Sloan Kettering Institute, Memorial Sloan Kettering Cancer Center, New York, NY}

\date{\today}


\begin{abstract}

Atomistic molecular simulations are a powerful way to make quantitative predictions, but the accuracy of these predictions depends entirely on the quality of the forcefield employed.
While experimental measurements of fundamental physical properties offer a straightforward approach for evaluating forcefield quality, the bulk of this information has been tied up in formats that are not machine-readable.
Compiling benchmark datasets of physical properties from non-machine-readable sources require substantial human effort and is prone to accumulation of human errors, hindering the development of reproducible benchmarks of forcefield accuracy.
Here, we examine the feasibility of benchmarking atomistic forcefields against the NIST ThermoML data archive of physicochemical measurements, which aggregates thousands of experimental measurements in a portable, machine-readable, self-annotating format. 
As a proof of concept, we present a detailed benchmark of the generalized Amber small molecule forcefield (GAFF) using the AM1-BCC charge model against measurements (specifically bulk liquid densities and static dielectric constants at ambient pressure) automatically extracted from the archive, and discuss the extent of available data.  
The results of this benchmark highlight a general problem with fixed-charge forcefields in the representation low dielectric environments such as those seen in binding cavities or biological membranes.  


\end{abstract}

\maketitle


\section{Introduction}

Recent advances in hardware and software for molecular dynamics simulation now permit routine access to atomistic simulations at the 100 ns timescale and beyond~\cite{salomon2013routine}.
Leveraging these advances in combination with consumer GPU clusters, distributed computing, or custom hardware has brought microsecond and millisecond simulation timescales within reach of many laboratories.  
These dramatic advances in sampling, however, have revealed deficiencies in forcefields as a critical barrier to enabling truly predictive simulations of physical properties of biomolecular systems.  

Protein and water forcefields have been the subject of numerous benchmarks~\cite{lindorff2012systematic, beauchamp2012protein, best2008} and enhancements~\cite{li2011iterative, best2012optimization, Lindorff-Larsen2010}, with key outcomes including the ability to fold fast-folding proteins \cite{shaw2011, ensign2007heterogeneity, Voelz2010}, improved fidelity of water thermodynamic properties~\cite{horn2004}, and improved prediction of NMR observables.  
Although small molecule forcefields have also been the subject of benchmarks~\cite{caleman2011force, fischer2015properties, zhang2015force} and improvements~\cite{fennell2014fixed}, such work has typically focused on small perturbations to specific functional groups.  
For example, a recent study found that modified hydroxyl nonbonded parameters led to improved prediction of static dielectric constants and hydration free energies~\cite{fennell2014fixed}.
There are also outstanding questions of generalizability of these targeted perturbations; it is uncertain whether changes to the parameters for a specific chemical moiety will be compatible with seemingly unrelated improvements to other groups.
Addressing these questions requires establishing community agreement upon shared benchmarks that can be easily replicated among laboratories to test proposed forcefield enhancements and expanded as the body of experimental data grows.

A key barrier to establishing reproducible and extensible forcefield accuracy benchmarks is that many experimental datasets are heterogeneous, paywalled, and unavailable in machine-readable formats (although notable counterexamples exist, e.g.~the RCSB~\cite{Berman2000}, FreeSolv~\cite{freesolv}, and the BMRB~\cite{Ulrich2008}).  
While this inconvenience is relatively minor for benchmarking forcefield accuracy for a single target system (e.g.~water), it becomes prohibitive for studies spanning the large relevant chemical spaces, such as forcefields intended to describe a large variety of druglike small organic molecules.  

In addition to inconvenience, the number and kind of human-induced errors that can corrupt hand-compiled benchmarks are legion.
A United States Geological Survey (USGS) case study examining the reporting and use of literature values of the aqueous solubility ($S_w$) and octanol-water partition coefficients ($K_{ow}$) for DDT and its persistent metabolite DDE provides incredible insight into a variety of common errors~\cite{usgs-ddt-report}.
Secondary sources are often cited as primary sources---a phenomenon that occurred up to five levels deep in the case of DDT/DDE;
citations for data are often incorrect, misattributed to unrelated publications, or omitted altogether;
numerical data can be mistranscribed, transposed, or incorrectly converted among unit systems~\cite{usgs-ddt-report}.
This occurs to such a degree that the authors note ``strings of erroneous data compose as much as 41--73 percent of the total data``~\cite{usgs-ddt-report}.
Given the incredible importance of these properties for human health and the environment, the quality of physicochemical datasets of far lesser importance is highly suspect.

To ameliorate problems of data archival, the NIST Thermodynamics Research Center (TRC) has developed an IUPAC standard XML-based format---ThermoML~\cite{frenkel2003thermoml, frenkel2006xml, chirico2013improvement}---for storing physicochemical measurements, uncertainties, and metadata.
Manuscripts containing new experimental measurements submitted to several journals (J.~Chem.~Eng.~Data, J.~Chem.~Therm., Fluid Phase Equil., Therm.~Acta, and Int.~J.~Therm.) are guided through a data archival process that involves sanity checks, conversion to a standard machine-readable format, and archival at the TRC (\url{http://trc.nist.gov/ThermoML.html}).  

Here, we examine the ThermoML archive as a potential source for a reproducible, extensible accuracy benchmark of biomolecular forcefields.
In particular, we concentrate on two important physical property measurements easily computable in many simulation codes---neat liquid density and static dielectric constant measurements---with the goal of developing a standard benchmark for validating these properties in fixed-charge forcefields of drug-like molecules and biopolymer residue analogues.  
These two properties provide sensitive tests of forcefield accuracy that are nonetheless straightforward to calculate.  
Using these data, we evaluate the generalized Amber small molecule forcefield (GAFF)~\cite{gaff,gaff2} with the AM1-BCC charge model~\cite{am1bcc1,am1bcc2} and identify systematic biases to aid further forcefield refinement.


\section{Methods}
\label{section:methods}

\subsection{ThermoML Archive retrieval and processing}
\label{section:thermoml-archive-retrieval}


A tarball archive snapshot of the ThermoML Archive was obtained from the the NIST TRC on 8 Apr. 2015.
To explore the content of this archive, we created a Python (version 2.7.9) tool (ThermoPyL: \url{https://github.com/choderalab/ThermoPyL}) that formats the XML content into a spreadsheet-like format accessible via the Pandas (version 0.15.2) library.  
First, we obtained the XML schema (\url{http://media.iupac.org/namespaces/ThermoML/ThermoML.xsd}) defining the layout of the data.
This schema was converted into a Python object via PyXB 1.2.4 (\url{http://pyxb.sourceforge.net/}).
Finally, this schema was used to extract the data into Pandas~\cite{pandas} dataframes, and the successive data filters described in Section~\ref{section:filtering-thermoml} were applied.  

\subsection{Simulation}
\label{section:simulation}

To enable automated accuracy benchmarking of physicochemical properties of neat liquids such as mass density and dielectric constant, we developed a semi-automated pipeline for preparing simulations, running them on a standard computer cluster using a portable simulation package, and analyzing the resulting data.
All code for this procedure is available at  \url{https://github.com/choderalab/LiquidBenchmark}.
Below, we describe the operation of the various stages of this pipeline and their application to the benchmark reported here.

\subsubsection{Preparation}
\label{section:preparation}

Chemical names were parsed from the ThermoML extract and converted to both CAS and smiles strings using cirpy \cite{swain2012cirpy}.  Smiles strings were converted into molecular structures using the OpenEye Python Toolkit version 2015-2-3 \cite{openeye}, as wrapped in openmoltools.  

Simulation boxes containing 1000 molecules were constructed using PackMol version 14-225~\cite{martinez2009packmol, packmolurl} wrapped in the Python automation library openmoltools.
In order to ensure stable automated equilibration, PackMol box volumes were chosen to accommodate twice volume of the enclosed atoms, with atomic radii estimated as 1.06 \AA\ and 1.53 \AA\ for hydrogens and nonhydrogens, respectively.  

For this illustrative benchmark, we utilized the generalized Amber small molecule forcefield (GAFF)~\cite{gaff,gaff2} with the AM1-BCC charge model~\cite{am1bcc1,am1bcc2}, which we shall refer to as the GAFF/AM1-BCC forcefield.

Canonical AM1-BCC~\cite{am1bcc1,am1bcc2, velez2014time} charges were generated with the OpenEye Python Toolkit version 2015-2-3 \cite{openeye}, using the {\tt oequacpac.OEAssignPartialCharges} module with the {\tt OECharges\_AM1BCCSym} option, which utilizes a conformational expansion procedure (using {\tt oeomega.OEOmega} \cite{hawkins2012conformer}) prior to charge fitting to minimize artifacts from intramolecular contacts.  
The {\tt OEOmega} selected conformer was then processed using {\tt antechamber} (with {\tt parmchk2}) and {\tt tleap} in AmberTools~14~\cite{amber14} to produce Amber-format {\tt prmtop} and {\tt inpcrd} files, which were then read into OpenMM to perform molecular simulations using the {\tt simtk.openmm.app} module.

The simulations reported here used libraries openmoltools~0.6.4~\cite{openmoltools}, OpenMM~6.3~\cite{eastman2012openmm}, and MDTraj~1.3~\cite{mcgibbon2014mdtraj}.  
Exact commands to install various dependencies can be found in Appendix~\ref{section:commands}.

\subsubsection{Equilibration and production}
\label{section:production}

Simulation boxes were first minimized using the L-BFGS algorithm \cite{liu1989limited} and equilibrated for $10^7$ steps with an equilibration timestep of 0.4 fs and a collision rate of 5 ps$^{-1}$.  
Production simulations were performed with OpenMM~6.3~\cite{eastman2012openmm} using a Langevin Leapfrog integrator~\cite{izaguirre-sweet-pande:psb:2010:langevin-leapfrog} (with collision rate 1 ps$^{-1}$) and a 1~fs timestep, as we found that timesteps of 2~fs timestep or greater led to a significant timestep dependence in computed equilibrium densities (Fig.~\ref{figure:timestep}).  

Equilibration and production simulations utilized a Monte Carlo barostat with a control pressure of 1 atm (101.325 kPa), utilizing molecular scaling and automated step size adjustment during equilibration, with volume moves attempted every 25 steps.  
The particle mesh Ewald (PME) method with conducting boundary conditions~\cite{Darden1993} was used with a long-range cutoff of 0.95~nm and a long-range isotropic dispersion correction. 
PME grid and spline parameters were automatically selected using the default settings in OpenMM 6.3 for the CUDA platform~\cite{eastman2012openmm}.

{\bf Automatic termination criteria.}
Production simulations were continued until automatic analysis showed standard errors in densities were less than $2 \times 10^{-4}$ g / $cm^{3}$.
Automatic analysis of the production simulation data was run every 1 ns of simulation time, and utilized the {\tt detectEquilibration} method in the timeseries module of pymbar 2.1~\cite{shirts2008statistically} to automatically discard the initial portion of the production simulation containing strong far-from-equilibrium behavior by maximizing the number of effectively uncorrelated samples in the remainder of the production simulation as determined by autocorrelation analysis using the fast adaptive statistical inefficiency computation method as implemented in the {\tt timeseries.computeStatisticalInefficiency} method of pymbar 2.1 (where the algorithm is described in~\cite{chodera2007}).
This approach is essentially the same as the fixed-width procedure described by eq. 7.12 of ref. \cite{brooks2011handbook}, with $n^*$ equal to 4000 and the sequential testing correction ($n^{-1}$ term) ignored due to the large value of $n$.
Statistical errors were computed by $\delta^2 \rho \approx \var{\rho} / N_\mathrm{eff}$, where $\var{\rho}$ is the sample variance of the density and $N_\mathrm{eff}$ is the number of effectively uncorrelated samples.  
With this protocol, we found starting trajectory lengths of $12000$ $(8000, 16000)$ frames (250 fs each), discarded regions of $28$  $(0, 460)$, and statistical inefficiencies of $20$ $(15, 28)$; reported numbers indicate (median, (25\%, 75\%)).  

Instantaneous densities were stored every 250~fs, while trajectory snapshots were stored every 5~ps.  

\subsection{Timings}

The wall time required for a given simulation depends on the number of atoms (3,000 - 29,000), the GPU used (GTX 680 or GTX Titan), and the time required for automated termination.  For butyl acrylate (21,000 atoms) on a GTX Titan, the wall-clock performance is approximately 80 ns / day.  Using 80 ns / day with approximately 3 ns of production simulation corresponds to 1 hour for the production segment of the simulation and 3 hours for the fixed equilibration portion of $10^7$ steps.  

\subsubsection{Data analysis and statistical error estimation}

Trajectory analysis was performed using OpenMM~6.3~\cite{eastman2012openmm} and MDTraj~1.3~\cite{mcgibbon2014mdtraj}.  

{\bf Mass density.}
Mass density $\rho$ was computed via the relation,
\begin{eqnarray}
\rho = \left\langle \frac{M}{V} \right\rangle \label{equation:mass-density} ,
 \end{eqnarray}
where $M$ is the total mass of all particles in the system and $V$ is the instantaneous volume of the simulation box.

{\bf Static dielectric constants.}
Static dielectric constants were calculated using the dipole fluctuation approach appropriate for PME with conducting (``tin-foil'') boundary conditions~\cite{horn2004, neumann1983dipole}, with the total system box dipole $\mu$ computed from trajectory snapshots using MDTraj 1.3~\cite{mcgibbon2014mdtraj}.
\begin{eqnarray}
\epsilon = 1 + \beta \frac{4\pi}{3} \frac{\langle \mu \cdot \mu \rangle - \langle \mu \rangle \cdot \langle \mu \rangle}{\langle V \rangle} \label{equation:dielectric_calculation}
\end{eqnarray}
where $\beta \equiv 1 / k_B T$ is the inverse temperature.

{\bf Computation of expectations.}
Expectations were estimated by computing sample means over the production simulation after discarding the initial far-from-equilibrium portion to equilibration (as described in {\bf Automatic termination criteria} above).

{\bf Statistical uncertainties.}
For density uncertainties, the Markov chain standard error (MCSE) was estimated as $\frac{\sigma}{\sqrt{N_{eff}}}$, where $\sigma$ is the density standard deviation of the simulation not discarded to equilibration, $N_{eff} = \frac{N}{g}$ is the effective sample size, and $g$ is the statistical inefficiency as estimated from the density time series.
For dielectric uncertainties, the portion of the production simulation not discarded to equilibration was used as input to a circular block bootstrapping procedure \cite{sheppard_2015_15681} with block sizes automatically selected to maximize the error \cite{flyvbjerg1989error}.

\subsubsection{Code availability}

Data analysis, all intermediate data (except configurational trajectories, due to their large size), and figure creation code for this work is available at \url{https://github.com/choderalab/LiquidBenchmark}.  


\section{Results}

\subsection{Extracting neat liquid measurements from the NIST TRC ThermoML Archive}
\label{section:filtering-thermoml}

As described in Section~\ref{section:thermoml-archive-retrieval}, we retrieved a copy of the ThermoML Archive and performed a number of sequential filtering steps to produce an ThermoML extract relevant for benchmarking forcefields describing small organic molecules.  
As our aim is to explore neat liquid data with functional groups relevant to biopolymers and drug-like molecules, we applied the following ordered filters, starting with all data containing density or static dielectric constants: 
\begin{enumerate}
 \item The measured sample contains only a single component (e.g.~no binary mixtures)
 \item The molecule contains only druglike elements (defined here as H, N, C, O, S, P, F, Cl, Br)
 \item The molecule has $\le$ 10 non-hydrogen atoms
 \item The measurement was performed in a biophysically relevant temperature range $(270 \le T$ [K] $\le 330)$
 \item The measurement was performed at ambient pressure $(100 \le P$ [kPa]  $\le 102)$
 \item Only measurements in liquid phase were retained
 \item The temperature and pressure were rounded to nearby values (as described below), averaging all measurements within each group of like conditions
 \item Only conditions (molecule, temperature, pressure) for which \emph{both} density and dielectric constants were available were retained
\end{enumerate}
The temperature and pressure rounding step was motivated by common data reporting variations; for example, an experiment performed at the freezing temperature of water and ambient pressure might be entered as either 101.325~kPa or 100~kPa, with a temperature of either 273~K or 273.15~K.  
Therefore all pressures within the range [kPa] $(100 \le P \le 102)$ were rounded to exactly 1 atm (101.325 kPa).  
Temperatures were rounded to one decimal place in K. 

The application of these filters (Table~\ref{table:ThermoMLSummary}) leaves 246 conditions---where a \emph{condition} here indicates a (molecule, temperature, pressure) tuple---for which both density and dielectric data are available.  
The functional groups present in the resulting dataset are summarized in Table~\ref{table:FunctionalGroups}; see Section~\ref{section:thermoml-archive-retrieval} for further description of the software pipeline used.  


\begin{table}
\begin{tabular}{lrr}
\hline
 &  \multicolumn{2}{c}{\bf Number of measurements remaining} \\ \cline{2-3}
{\bf Filter step} &  {\bf Mass density} &  {\bf Static dielectric} \\ 
\hline
1.  Single Component   &               136212 &                                     1651 \\
2.  Druglike Elements  &               125953 &                                     1651 \\
3.  Heavy Atoms        &                71595 &                                     1569 \\
4.  Temperature        &                38821 &                                      964 \\
5.  Pressure           &                14103 &                                      461 \\
6.  Liquid state       &                14033 &                                      461 \\
7.  Aggregate T, P     &                 3592 &                                      432 \\
8.  Density+Dielectric &                  246 &                                      246 \\

\hline
\end{tabular}
\caption{{\bf Successive filtration of the ThermoML Archive.}
A set of successive filters were applied to all measurements in the ThermoML Archive that contained either mass density or static dielectric constant measurements.
Each column reports the number of measurements remaining after successive application of the corresponding filtration step.  
The 246 final measurements correspond to 45 unique molecules measured at several temperature conditions.  
}
\label{table:ThermoMLSummary}
\end{table}

\begin{table}
\begin{tabular}{lr}
\toprule
{\bf Functional Group} &    {\bf Occurrences} \\
\midrule
1,2-aminoalcohol                                    &   4 \\
1,2-diol                                            &   3 \\
alkene                                              &   3 \\
aromatic compound                                   &   1 \\
carbonic acid diester                               &   2 \\
carboxylic acid ester                               &   4 \\
dialkyl ether                                       &   7 \\
heterocyclic compound                               &   3 \\
ketone                                              &   3 \\
lactone                                             &   1 \\
primary alcohol                                     &  19 \\
primary aliphatic amine (alkylamine)                &   2 \\
primary amine                                       &   2 \\
secondary alcohol                                   &   4 \\
secondary aliphatic amine (dialkylamine)            &   2 \\
secondary aliphatic/aromatic amine (alkylarylamine) &   1 \\
secondary amine                                     &   3 \\
sulfone                                             &   1 \\
sulfoxide                                           &   1 \\
tertiary aliphatic amine (trialkylamine)            &   3 \\
tertiary amine                                      &   3 \\
\bottomrule
\end{tabular}
\caption{{\bf Functional groups present in filtered dataset.}  
The filtered ThermoML dataset contained 246 distinct (molecule, temperature, pressure) conditions, spanning 45 unique compounds.
The functional groups represented in these compounds (as identified by the program {\tt checkmol} v0.5~\cite{haider2010functionality}) is summarized here.
}
\label{table:FunctionalGroups}
\end{table}


\subsection{Benchmarking GAFF/AM1-BCC against the ThermoML Archive}

\subsubsection{Mass density}

\begin{figure}

\subfigure[]{
\includegraphics[width=\columnwidth]{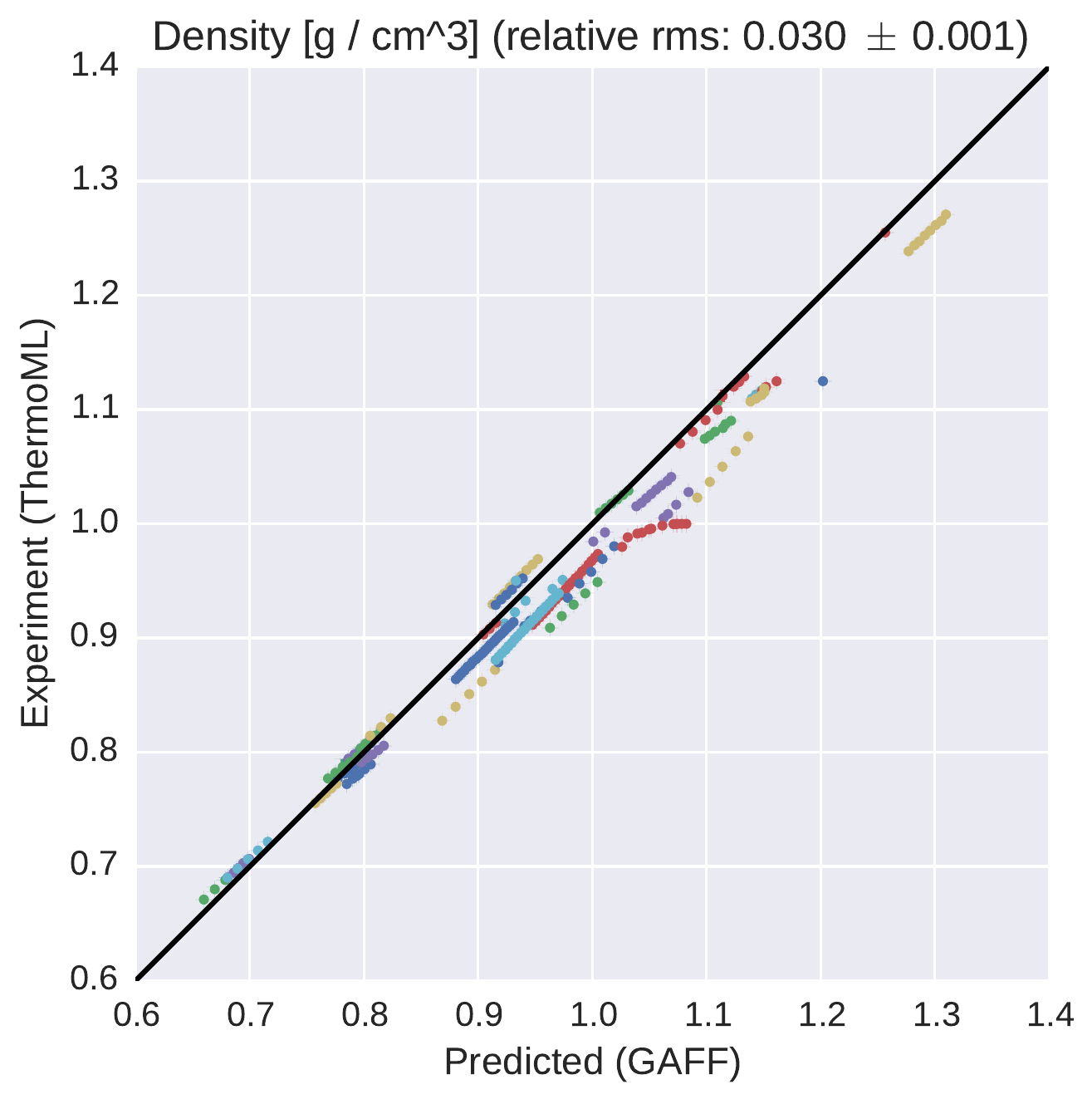}
}

\subfigure[]{
\includegraphics[width=\columnwidth]{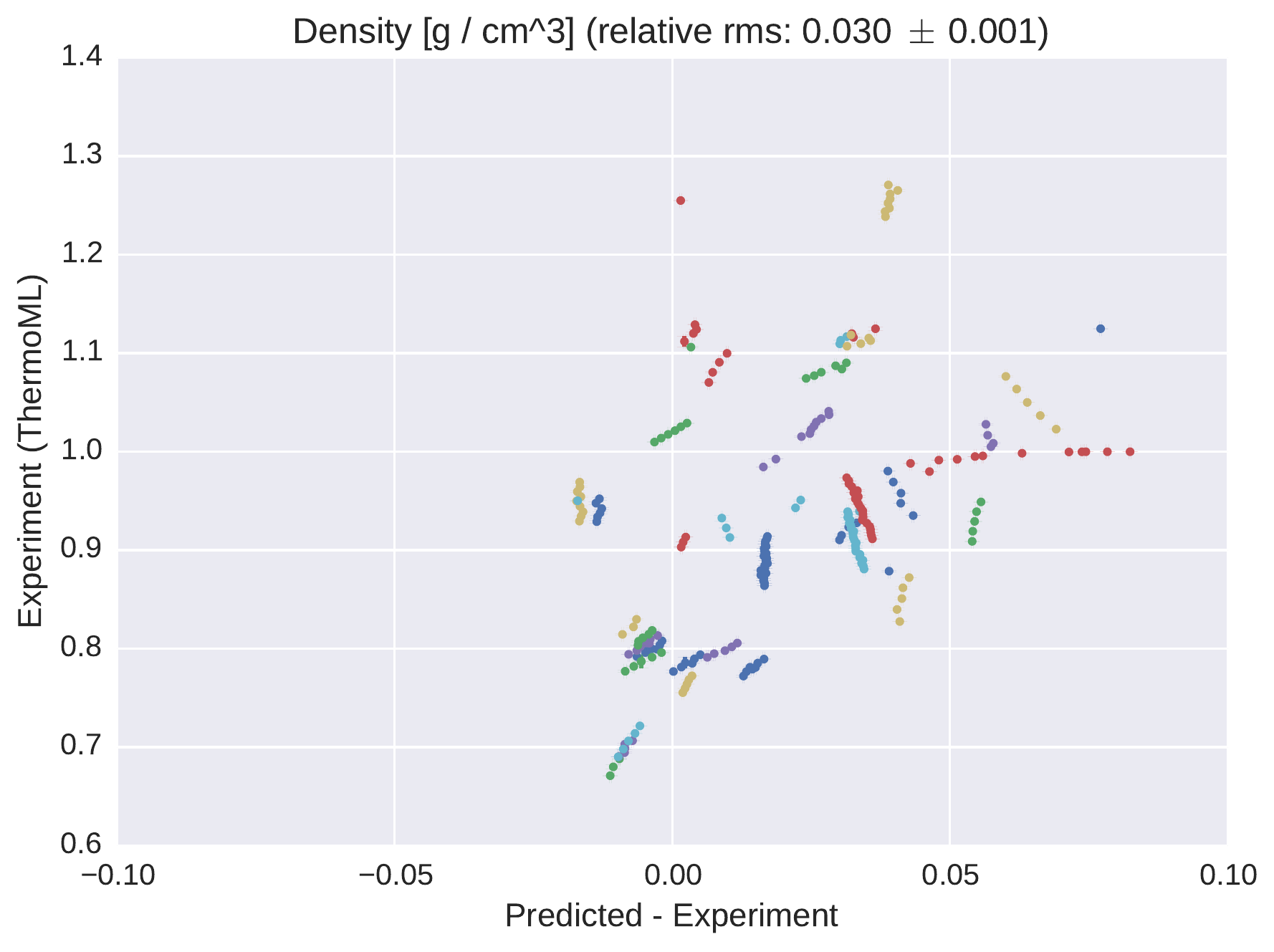}
}

\caption{{\bf Comparison of liquid densities between experiment and simulation.}
(a).  Liquid density measurements extracted from ThermoML are compared against densities predicted using the GAFF / AM1-BCC small molecule fixed-charge forcefield.
Color groupings represent identical chemical species, although the color map repeats itself due to the large (45) number of unique compounds.
Plots of density versus temperature grouped by chemical species are available in Fig.~\ref{figure:AllDensities}.
Simulation error bars represent one standard error of the mean, with the number of effective (uncorrelated) samples estimated using pymbar.  
Experimental error bars indicate the standard deviation between independently reported measurements, when available, or author-reported standard deviations in ThermoML entries; for some measurements, neither uncertainty estimate is available.  
See Fig.~\ref{figure:ErrorAnalysisDensity} for further discussion of error.  (b).  The same plot, but with the residual (predicted minus experiment) on the x axis.  Note that the error bars are all smaller than the symbols.  
}
\label{figure:Density}
\end{figure}

Mass densities of bulk liquids have been widely used for parameterizing and testing forcefields, particularly the Lennard-Jones parameters representing dispersive and repulsive interactions~\cite{jorgensen1983comparison, jorgensen1984optimized}.
We therefore used the present ThermoML extract as a benchmark of the GAFF/AM1-BCC forcefield (Fig.~\ref{figure:Density}).  

{\bf Overall accuracy.}
Overall, the densities show reasonable accuracy, with a root-mean square (RMS) relative error over all measurements of (3.0$\pm$0.1)\%, especially encouraging given that this forcefield was not designed with the intention of modeling bulk liquid properties of organic molecules~\cite{gaff,gaff2}.
This is reasonably consistent with previous studies reporting agreement of 4\% on a different benchmark set~\cite{caleman2011force}.

{\bf Temperature dependence.}
For a given compound, the signs of the errors typically do not change at different temperatures (Fig.~\ref{figure:Density}, Fig.~\ref{figure:AllDensities}).  
Furthermore, the magnitudes of the error also remain largely constant (vertical lines in Fig.~\ref{figure:Density} B), although several exceptions do occur.  
It is possible that these systematic density offsets indicate correctable biases in forcefield parameters.  


{\bf Outliers.}
The largest density errors occur for a number of oxygen-containing compounds: 1,4-dioxane; 2,5,8-trioxanonane; 2-aminoethanol; dimethyl carbonate; formamide; and water (Fig.~\ref{figure:AllDensities}).
The absolute error on these poor predictions is on the order of 0.05  g/$cm^{3}$, which is substantially higher than the measurement error ($\le0.008$ g/$cm^{3}$; see Fig.~\ref{figure:ErrorAnalysisDensity}).  

We note that our benchmark includes a GAFF/AM1-BCC model for water due to our desire to automate benchmarks against a forcefield capable of modeling a large variety of small molecular liquids.
Water---an incredibly important solvent in biomolecular systems---is generally treated with a special-purpose model (such as TIP3P~\cite{jorgensen1983comparison} or TIP4P-Ew~\cite{horn2004}) parameterized to fit a large quantity of thermophysical data.
As expected, the GAFF/AM1-BCC model performs poorly in reproducing liquid densities for this very special solvent.
We conclude that it remains highly advisable that the field continue to use specialized water models when possible.


\subsubsection{Static dielectric constant}
\label{section:results:static-dielectric-constant}

\begin{figure}
\includegraphics[width=\columnwidth]{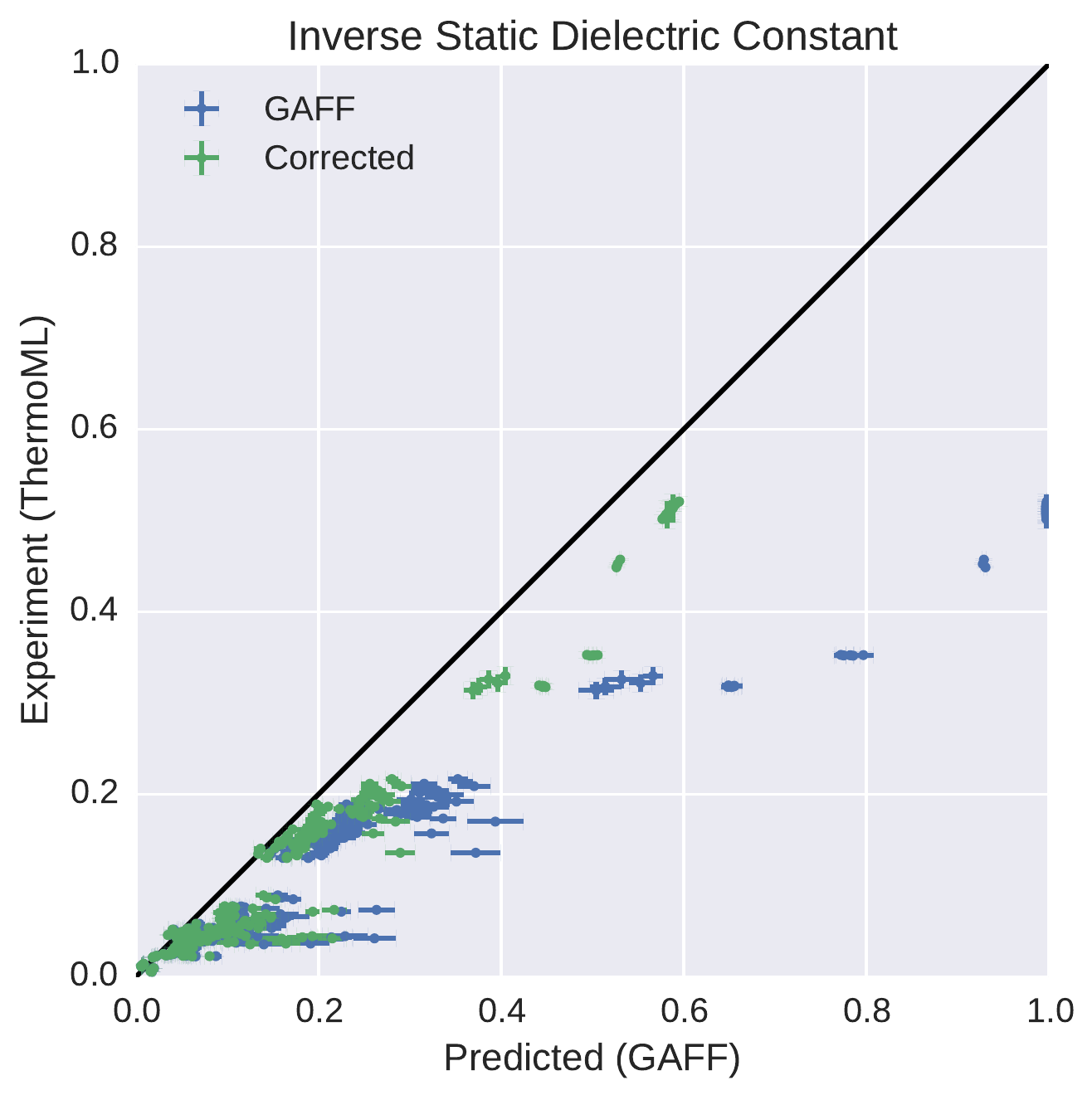}

\caption{{\bf Measured (ThermoML) versus predicted (GAFF / AM1-BCC) inverse static dielectrics (a).}
Simulation error bars represent one standard error of the mean.  
Experimental error bars indicate the larger of standard deviation between independently reported measurements and the authors reported standard deviations; for some measurements, neither uncertainty estimate is available.  
See Fig.~\ref{figure:ErrorAnalysisDensity} for further discussion of error.  
See Section~\ref{section:results:static-dielectric-constant} for explanation of why inverse dielectric constant (rather than dielectric constant) is plotted. 
For nonpolar liquids, it is clear that the forcefield predicts electrostatic interactions that are substantially biased by missing polarizability.
Plots of dielectric constant versus temperature grouped by chemical species are available in Fig.~\ref{figure:AllDielectrics}.
}
\label{figure:Dielectric}
\end{figure}

{\bf Overall accuracy.}
As a measure of the dielectric response, the static dielectric constant of neat liquids provides a critical benchmark of the accuracy electrostatic treatment in forcefield models.  
Discussing the accuracy in terms the ability of GAFF/AM1-BCC to reproduce the static dielectric constant $\epsilon$ is not necessarily meaningful because of the way that the solvent dielectric $\epsilon$ enters into the Coulomb potential between two point charges separated by a distance $r$,
\begin{eqnarray}
U(r) = \frac{q_1 q_2}{\epsilon \, r} \propto \frac{1}{\epsilon} .
\end{eqnarray}
It is evident that $1/\epsilon$ is a much more meaningful quantity to compare than $\epsilon$ directly, as a 5\% error in $1/\epsilon$ will cause a 5\% error in the Coulomb potential between two point charges (assuming a uniform dielectric), while a 5\% error in $\epsilon$ will have a much more complex $\epsilon$-dependent effect on the Coulomb potential.
We therefore compare simulations against measurements in our ThermoML extract on the $1/\epsilon$ scale in Fig.~\ref{figure:Dielectric}.  


{\bf GAFF/AM1-BCC systematically underestimates the dielectric constants of nonpolar liquids.}
Overall, we find the dielectric constants to be qualitatively reasonable, but with clear deviations from experiment particularly for nonpolar liquids. 
This is not surprising given the complete neglect of electronic polarization which will be the dominant contribution for such liquids.
In particular, GAFF/AM1-BCC systematically underestimates the dielectric constants for nonpolar liquids, with the predictions of $\epsilon \approx 1.0 $ being substantially smaller than the measured $\epsilon \approx 2$.  
Because this deviation likely stems from the lack of an explicit treatment of electronic polarization, we used a simple empirical polarization model that computes the molecular electronic polarizability $\alpha$ as a sum of elemental atomic polarizability contributions~\cite{bosque2002polarizabilities}.


From the computed molecular electronic polarizability $\alpha$, an additive correction to the simulation-derived static dielectric constant accounting for the missing electronic polarizability can be computed~\cite{horn2004}
\begin{eqnarray}
\Delta \epsilon &=& 4 \pi N  \frac{\alpha}{\langle V \rangle} \label{equation:dielectric correction}
\end{eqnarray}
A similar polarization correction was used in the development of the TIP4P-Ew water model, where it had a minor effect~\cite{horn2004} because almost all the high static dielectric constant for water comes from the configurational response of its strong dipole.  
However, the missing polarizability is a dominant contribution to the static dielectric constant of nonpolar organic molecules;   
in the case of water, the empirical atomic polarizability model predicts a dielectric correction of 0.52, while 0.79 was used for the TIP4P-Ew model.  
Averaging all liquids in the present work leads to polarizability corrections to the static dielectric of $0.74 \pm 0.08$.
Taking the dataset as a whole, we find that the relative error in uncorrected dielectric is on the order of $-0.34 \pm 0.02$, as compared to $-0.25 \pm 0.02$ for the corrected dielectric.


\section{Discussion}

\subsection{Mass densities}

Our simulations have indicated the presence of systematic density biases with magnitudes larger than the measurement error.  
Correcting these errors may be a low-hanging fruit for future forcefield refinements.
As an example of the feasibility of improved accuracy in densities, a recent three-point water model was able to recapitulate water density with errors of less than 0.005 g / $cm^{3}$ over temperature range [280 K, 320 K] \cite{wang2014building}.
This improved accuracy in density prediction was obtained alongside accurate predictions of other experimental observables, including static dielectric constant.  
We suspect that such accuracy might be obtainable for GAFF-like forcefields across some portion of chemical space.  
A key challenge for the field is to demarcate the fundamental limit of fixed-charge forcefields for predicting orthogonal classes of experimental observables.
For example, is it possible to achieve a relative density error of $10^{-4}$ without sacrificing accuracy of other properties such as enthalpies?
In our opinion, the best way to answer such questions is to systematically build forcefields with the goal of predicting various properties to within their known experimental uncertainties, similar to what has been done for water \cite{horn2004, wang2014building}.


\subsection{Dielectric constants in forcefield parameterization}

A key feature of the static dielectric constant for a liquid is that, for forcefield purposes, it consists of two very different components, distinguished by the dependence on the fixed charges of the forcefield and dynamic motion of the molecule. One component, the high-frequency dielectric constant, arises from the almost-instantaneous electronic polarization in response to the external electric field: this contributes a small component, generally around $\epsilon = 2$, which can be dominant for non-polar liquids but is completely neglected by the non-polarizable forcefields in common use for biomolecular simulations. The other component arises from the dynamical response of the molecule, through nuclear motion, to allow its various molecular multipoles to respond to the external electric field: for polar liquids such as water, this contributes the majority of the dielectric constant. Thus for polar liquids, we expect the parameterized atomic charges to play a major role in the static dielectric.  

Recent forcefield development has seen a resurgence of papers fitting dielectric constants during forcefield parameterization \cite{wang2014building, fennell2014fixed}.  
However, a number of authors have pointed out potential challenges in constructing self-consistent fixed-charge forcefields \cite{fennell2012simple, leontyev2014polarizable}.

\begin{figure}

\subfigure[]{
\includegraphics[width=\columnwidth]{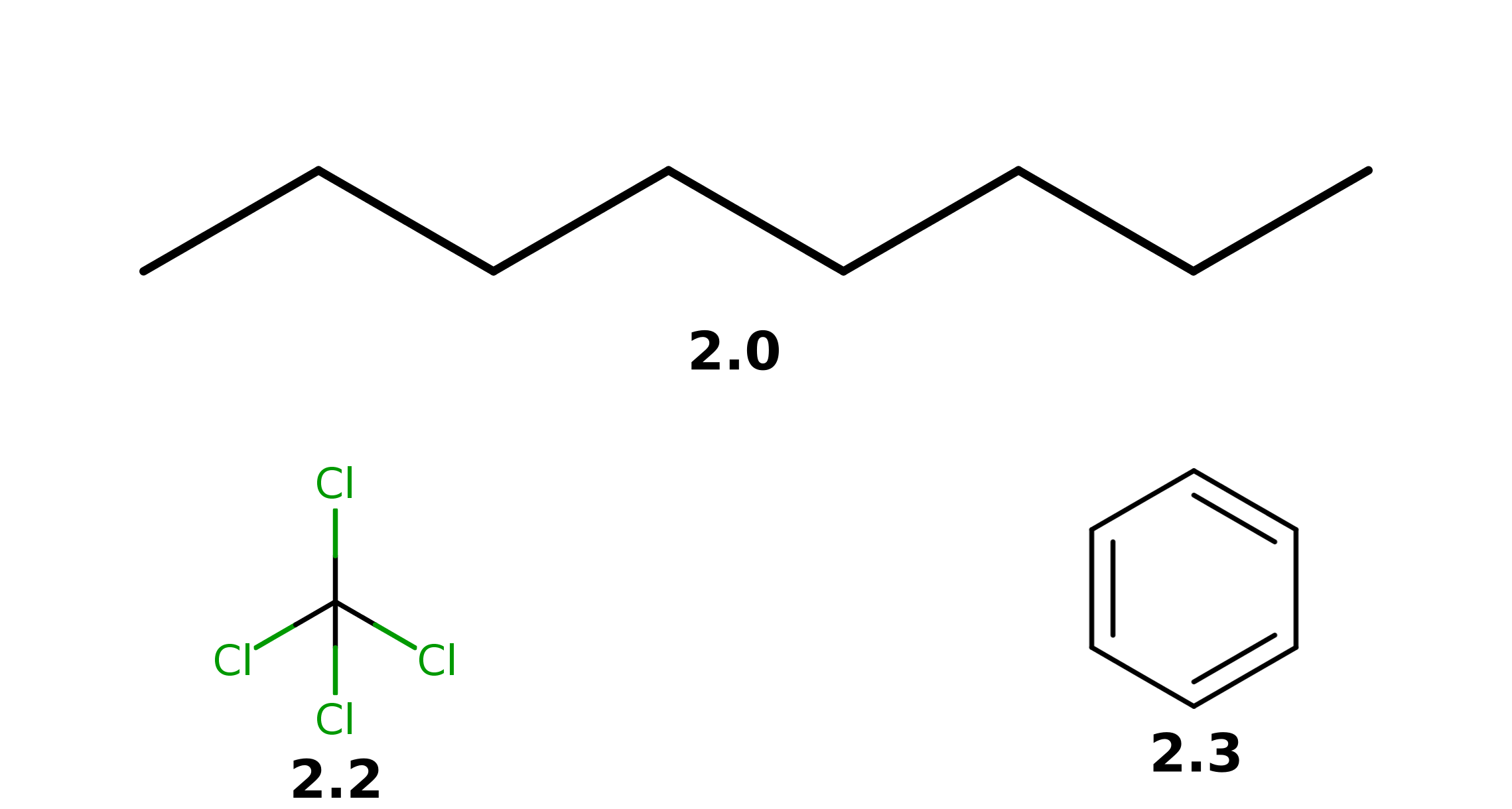}
}

\noindent\rule{8cm}{0.4pt}
\subfigure[]{
\includegraphics[width=\columnwidth]{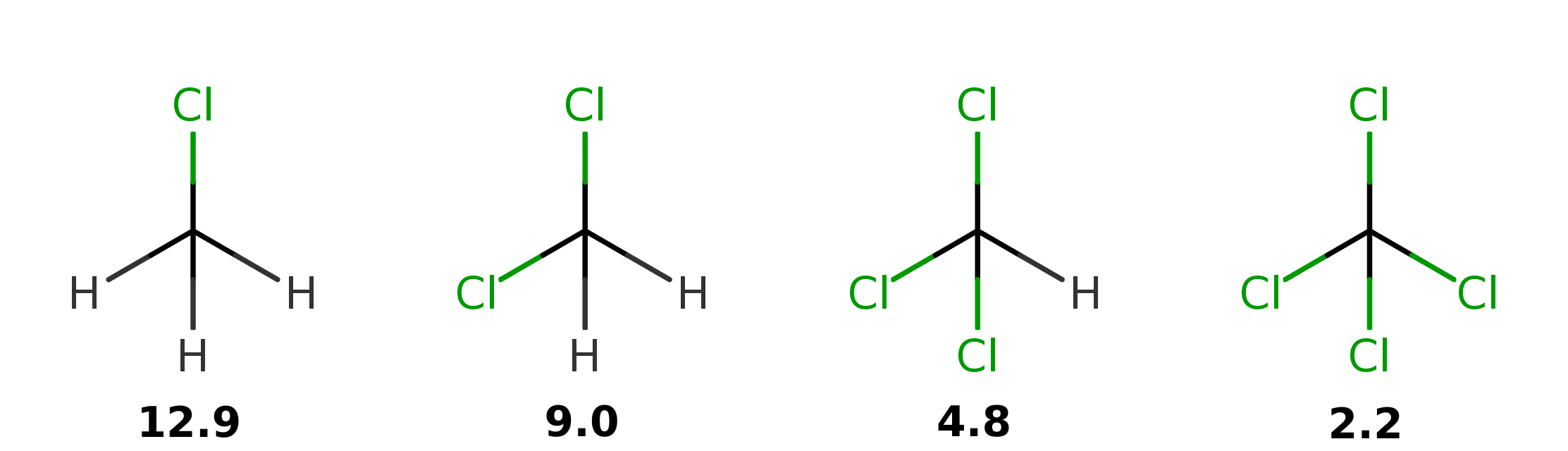}
}

\caption{{\bf Typical experimental static dielectric constants of some nonpolar compounds.}
(a). Measured static dielectric constants of various nonpolar or symmetric molecules~\cite{d1990dielectric, haynes2011crc}.  
Fixed-charge forcefields give $\epsilon \approx 1$ for each species; for example, we calculated $\epsilon = 1.0030 \pm 0.0002$ for octane.
(b).  A congeneric series of chloro-substituted methanes have static dielectric constants between 2 and 13.  
Reported dielectric constants are at near-ambient temperatures.  
}
\label{figure:nonpolars}

\end{figure}

Interestingly, recent work by Dill and coworkers~\cite{fennell2012simple} observed that, for $\mathrm{CCl_4}$, reasonable choices of point charges are incapable of recapitulating the observed dielectric of $\epsilon = 2.2$, instead producing dielectric constants in the range of $1.0 \le \epsilon \le 1.05$.  
This behavior is quite general: fixed point charge forcefields will predict $\epsilon \approx 1$ for many nonpolar or symmetric molecules, but the measured dielectric constants are instead $\epsilon \approx 2$ (Fig.~\ref{figure:nonpolars}).  
While this behavior is well-known and results from missing physics of polarizability, we suspect it may have several profound consequences, which we discuss below.


Suppose, for example, that one attempts to fit forcefield parameters to match the static dielectric constants of $\mathrm{CCl_4}$, $\mathrm{CHCl_3}$, $\mathrm{CH_2Cl_2}$, and $\mathrm{CH_3Cl}$.
In moving from the tetrahedrally-symmetric $\mathrm{CCl_4}$ to the asymmetric $\mathrm{CHCl_3}$, it suddenly becomes possible to achieve the observed dielectric constant of 4.8 by an appropriate choice of point charges.
However, the model for $\mathrm{CHCl_3}$ uses fixed point charges to account for \emph{both} the permanent dipole moment and the electronic polarizability, whereas the $\mathrm{CCl_4}$ model contains no treatment of polarizability.  
We hypothesize that this inconsistency in parameterization may lead to strange mismatches, where symmetric molecules (e.g.~benzene and $\mathrm{CCl_4}$) have qualitatively different properties than closely related asymmetric molecules (e.g.~toluene and $\mathrm{CHCl_3}$).

How important is this effect?
We expect it to be important wherever we encounter the transfer of a polar molecule (such as a peptide, native ligand, or a pharmaceutical small molecule) from a polar environment (such as the cytosol, interstitial fluid, or blood) into a non-polar environment (such as a biological membrane or non-polar binding site of an enzyme or receptor). Thus we expect this to be implicated in biological processes ranging from ligand binding to absorption and distribution within the body. To understand this conceptually, consider the transfer of a polar small-molecule transfer from the non-polar interior of a lipid bilayer to the aqueous and hence very polar cytosol.
As a possible real-world example, we imagine that the missing atomic polarizability could be important in accurate transfer free energies involving low-dielectric solvents, such as the small-molecule transfer free energy from octanol or cyclohexane to water.  
The Onsager model for solvation of a dipole $\mu$ of radius $a$ gives us a way to estimate the magnitude of error introduced by making an error of $\Delta \epsilon$ in the static dielectric constant of a solvent.
The free energy of dipole solvation is given by this model as
\begin{equation} \label{eq:onsager}
\Delta G = -\frac{\mu^2}{a^3}\frac{\epsilon - 1}{2 \epsilon + 1}
\end{equation}
such that, for an error of $\Delta \epsilon$ departing from the true static dielectric constant $\epsilon$, we find the error in solvation is
\begin{equation} \label{equation:onsager-error}
\Delta \Delta G = -\frac{\mu^2}{a^3} \left[ \frac{(\epsilon+\Delta \epsilon) - 1}{2 (\epsilon+\Delta \epsilon) + 1} - \frac{\epsilon - 1}{2 \epsilon + 1} \right] 
\end{equation}
For example, the solvation of water ($a = 1.93$~\AA, $\mu = 2.2$~D) in a low dielectric medium such as tetrachloromethane or benzene ($\epsilon \sim 2.2$, but $\Delta \epsilon = -1.2$) gives an error of $\Delta \Delta G \sim -8$ kJ/mol (-2 kcal/mol).

{\bf Implications for transfer free energies.}
As another example, consider the transfer of small druglike molecules from a nonpolar solvent (such as cyclohexane) to water, a property often measured to indicate the expected degree of lipophilicity of a compound.
To estimate the magnitude of error expected, for each molecule in the latest (Feb. 20) FreeSolv database~\cite{freesolv, freesolv_github}, we estimated the expected error in computed transfer free energies should GAFF/AM1-BCC be used to model the nonpolar solvent cyclohexane using the Onsager model (Eq.~\ref{equation:onsager-error}).
We used took the cavity radius $a$ to be the half the maximum interatomic distance and calculated $\mu = \sum_i q_i r_i$ using the provided mol2 coordinates and AM1-BCC charges.  
This calculation predicts a mean error of ($-3.8 \pm 0.3$) kJ / mol [($-0.91 \pm0.07$) kcal / mol] for the 643 molecules (where the standard error is computed from bootstrapping over FreeSolv compound measurements), 
suggesting that the missing atomic polarizabilty unrepresentable by fixed point charge forcefields could contribute substantially to errors in predicted transfer and solvation properties of druglike molecules.  
In other words, the use of a fixed-charge physics may lead to errors of $3.8$ kJ / mol in cyclohexane transfer free energies.  
We conjecture that this missing physics will be important in the upcoming (2015) SAMPL challenge \cite{newman2009practical}, which will examine transfer free energies in several low dielectric media. 

{\bf Utility in parameterization.}
Given their ease of measurement and direct connection to long-range electrostatic interactions, static dielectric constants have high potential utility as primary data for forcefield parameterization efforts.  
Although this will require the use of forcefields with explicit treatment of atomic polarizability, the inconsistency of fixed-charge models in low-dielectric media is sufficiently alarming to motivate further study of polarizable forcefields.  In particular, continuum methods~\cite{truchon2010using, truchon2009integrated, truchon2008accurate}, point dipole methods~\cite{Ponder2010, ren2004temperature}, and Drude methods~\cite{lamoureux2003modeling, anisimov2005determination} have been maturing rapidly.  Finding the optimal balance of accuracy and performance remains an open question; however, the use of experimentally-parameterized direct polarization methods~\cite{wang2013systematic} may provide polarizability physics at a cost not much greater than fixed charge forcefields.

\subsection{ThermoML as a data source}

The present work has focused on the neat liquid density and dielectric measurements present in the ThermoML Archive~\cite{frenkel2006xml, frenkel2003thermoml, chirico2003thermoml} as a target for molecular dynamics forcefield validation.  
While liquid mass densities and static dielectric constants have already been widely used in forcefield work, several aspects of ThermoML make it a unique resource for the forcefield community.  
First, the aggregation, support, and dissemination of ThermoML datasets through the ThermoML Archive is supported by NIST, whose mission makes these tasks a long-term priority.  
Second, the ThermoML Archive is actively growing, through partnerships with several journals, and new experimental measurements published in these journals are critically examined by the TRC and included in the archive.  
Finally, the files in the ThermoML Archive are portable and machine readable via a formal XML schema, allowing facile access to hundreds of thousands of measurements.  
Numerous additional physical properties contained in ThermoML---including activity coefficients, diffusion constants, boiling point temperatures, critical pressures and densities, coefficients of expansion, speed of sound measurements, viscosities, excess molar enthalpies, heat capacities, and volumes---for neat phases and mixtures represent a rich dataset of high utility for forcefield validation and parameterization.


\section{Conclusions}

High quality, machine-readable datasets of physicochemical measurements will be required for the construction of next-generation small molecule forcefields.  
Here we have discussed the NIST/TRC ThermoML archive as a growing source of physicochemical measurements that may be useful for the forcefield community.
From the NIST/TRC ThermoML archive, we selected a dataset of 246 ambient, neat liquid systems for which both densities and static dielectric constants are available.  
Using this dataset, we benchmarked GAFF/AM1-BCC, one of the most popular small molecule forcefields.
We find systematic biases in densities and particularly static dielectric constants.
Element-based empirical polarizabilty models are able to account for much of the systematic differences between GAFF/AM1-BCC and experiment.  
Non-polarizable forcefields may show unacceptable biases when predicting transfer and binding properties of non-polar environments such as binding cavities or membranes.  


\section{Acknowledgements}

We thank Patrick B.~Grinaway (MSKCC), Vijay S.~Pande (Stanford University), Lee-Ping Wang (Stanford University), Peter Eastman (Stanford University), Robert McGibbon (Stanford University), Jason Swails (Rutgers University), David L.~Mobley (University of California, Irvine), Michael R. Shirts (University of Virginia), William C. Swope (IBM), Julia E. Rice (IBM), Hans Horn (IBM), Jed W. Pitera (IBM), and members of Chodera lab for helpful discussions.  
Support for JMB was provided by the Tri-Institutional Training Program in Computational Biology and Medicine (via NIH training grant 1T32GM083937).  KAB was supported in part by Starr Foundation grant I8-A8-058.  JDC and KAB acknowledge partial support from NIH grant P30 CA008748.  KAB, JLB, ASR, and JDC acknowledge the generous support of this research by the Sloan Kettering Institute.


\section{Disclaimers}

This contribution of the National Institute of Standards and Technology (NIST) is not subject to copyright in the United States.  
Products or companies named here are cited only in the interest of complete technical description, and neither constitute nor imply endorsement by NIST or by the U.S.~government.  
Other products may be found to serve as well.

\clearpage

\section{TOC Figure}

\includegraphics[width=\columnwidth]{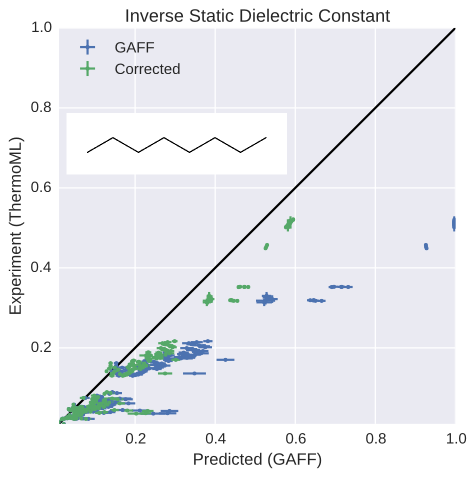}

\clearpage

\appendix 

\section{Appendices}

\begin{itemize}
 \item Figure: Timestep-dependence of density
 \item Figure: Error analysis (density) for ThermoML dataset
 \item Figure: Error analysis (static dielectric constant) for ThermoML dataset
 \item Figure: Temperature Dependence: Density
 \item Figure: Temperature Dependence: Static Dielectric Constant
 \item Commands to install dependencies
\end{itemize}

\clearpage

\begin{figure}

\subfigure[]{
\includegraphics[width=8cm]{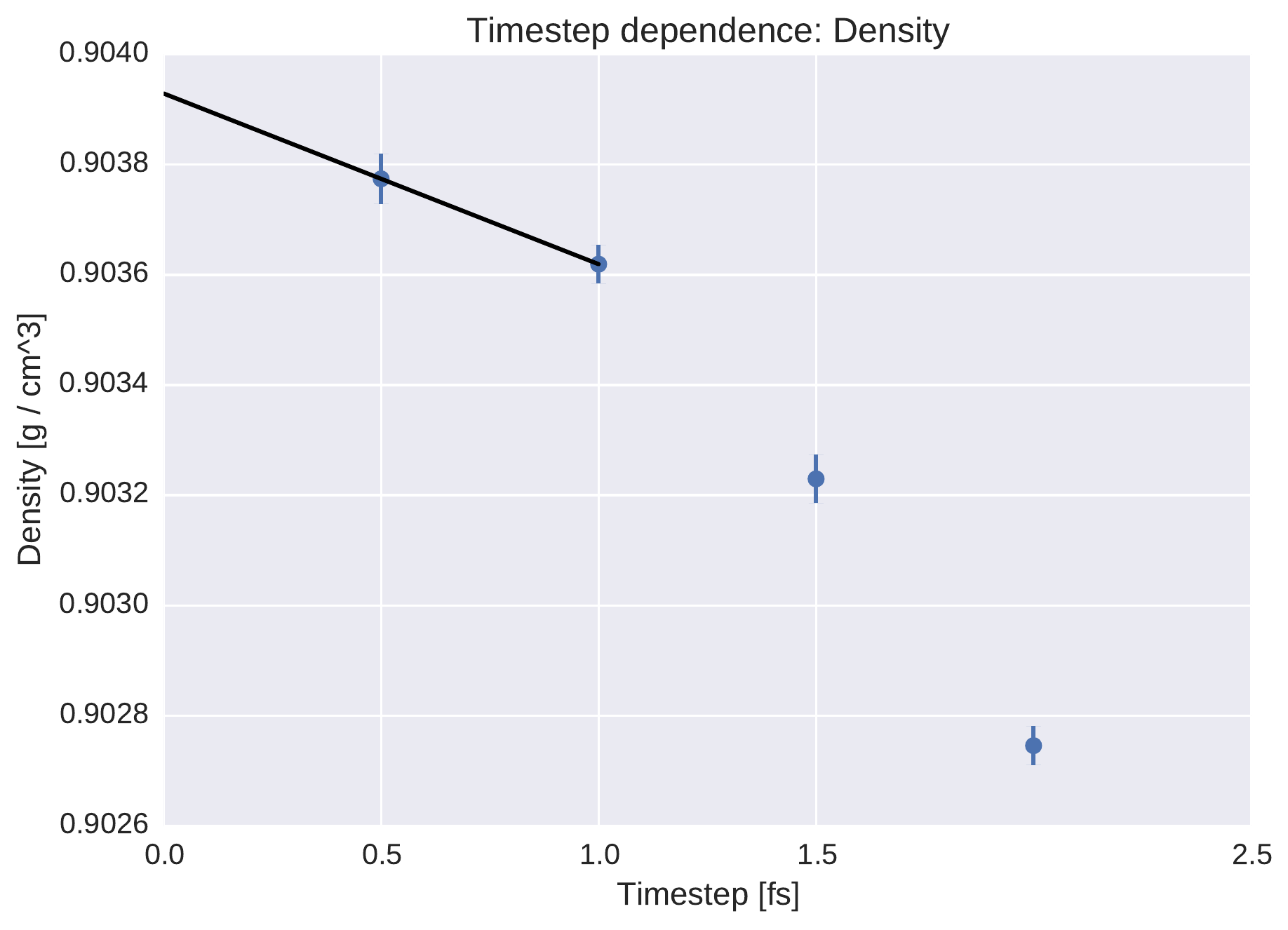}
}
\subfigure[]{
\includegraphics[width=8cm]{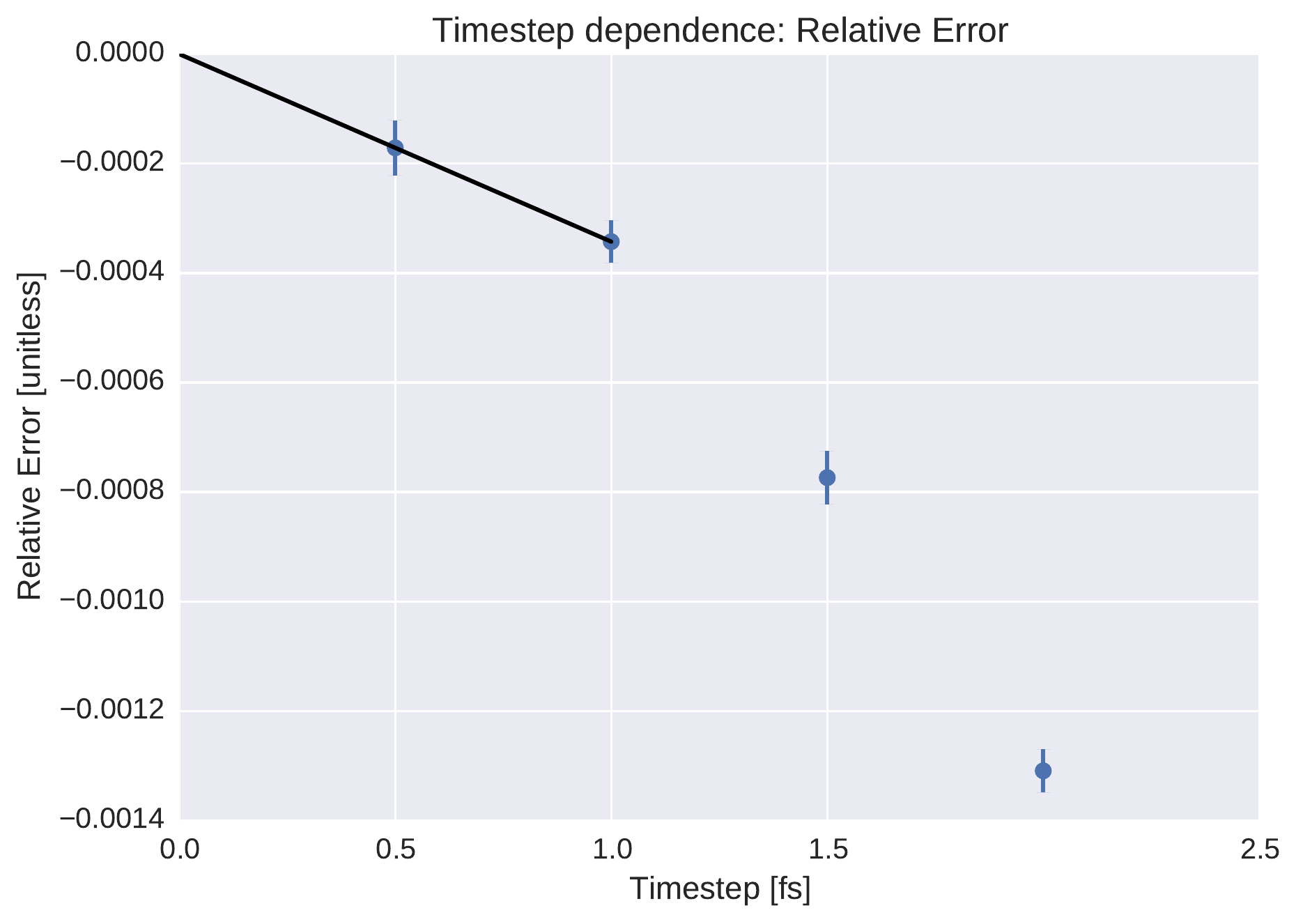}
}

\caption{
{\bf Dependence of computed density on simulation timestep.}
To probe the systematic error from finite time-step integration, we examined the timestep dependence of butyl acrylate density.  
(a).  The density is shown for several choices of timestep.  
(b).  The relative error, as compared to the reference value, is shown for several choices of timestep.  
Error bars represent standard errors of the mean, with the number of effective samples estimated using pymbar's statistical inefficiency routine \cite{shirts2008statistically}.  
The reference value is estimated by linear extrapolation to 0 fs using the 0.5 fs and 1.0 fs data points; the linear extrapolation is shown as black lines.  
We find a 2~fs timestep leads to systematic biases in the density on the order of 0.13\%, while 1 fs reduces the systematic bias to approximately 0.8\%---we therefore selected a 1~fs timestep for the present work, where we aimed to achieve three digits of accuracy in density predictions.
}
\label{figure:timestep}

\end{figure}



\clearpage

\begin{figure}

\subfigure[]{
\includegraphics[width=7.3cm]{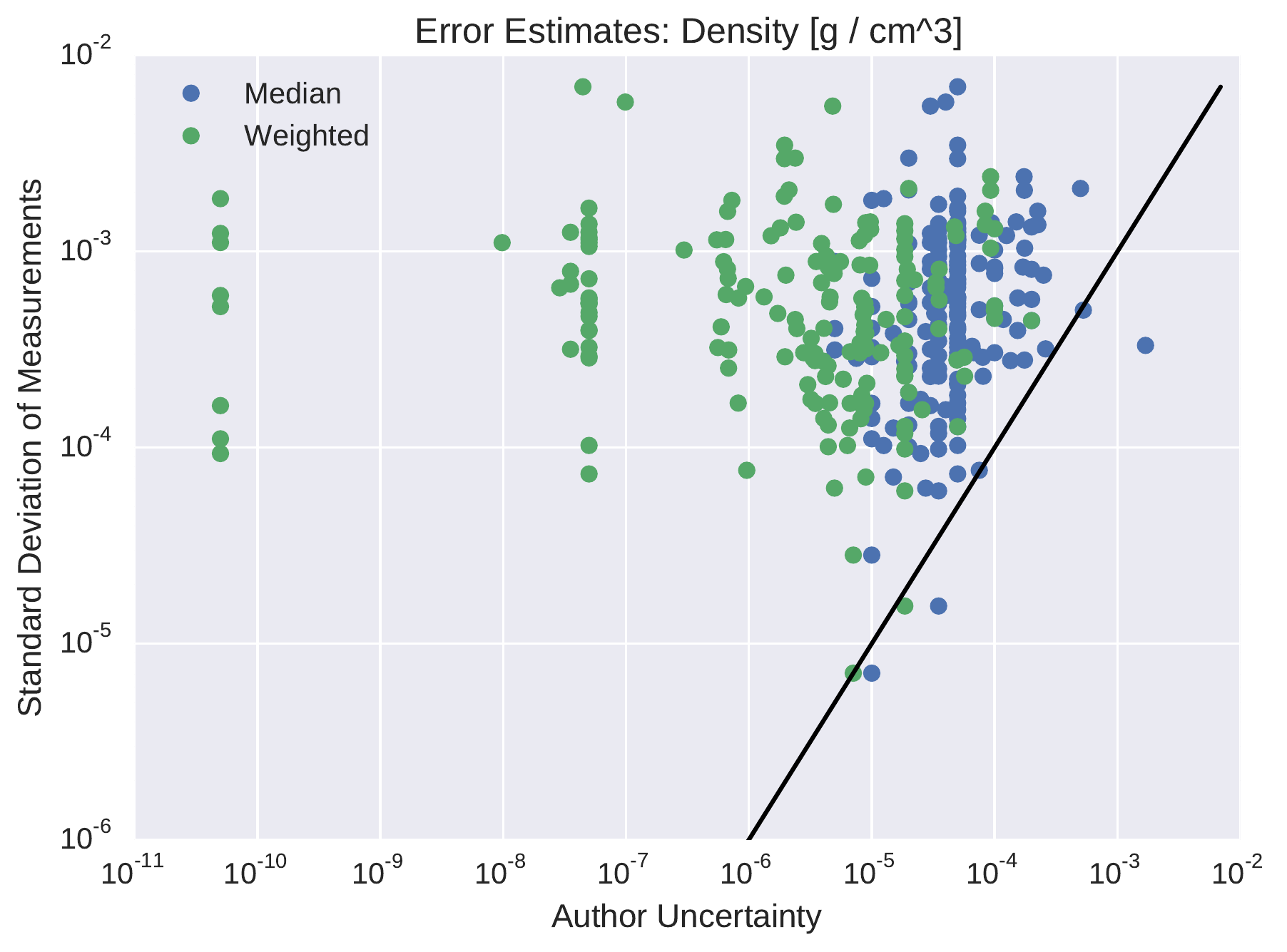}
}

\subfigure[]{
\includegraphics[width=7.3cm]{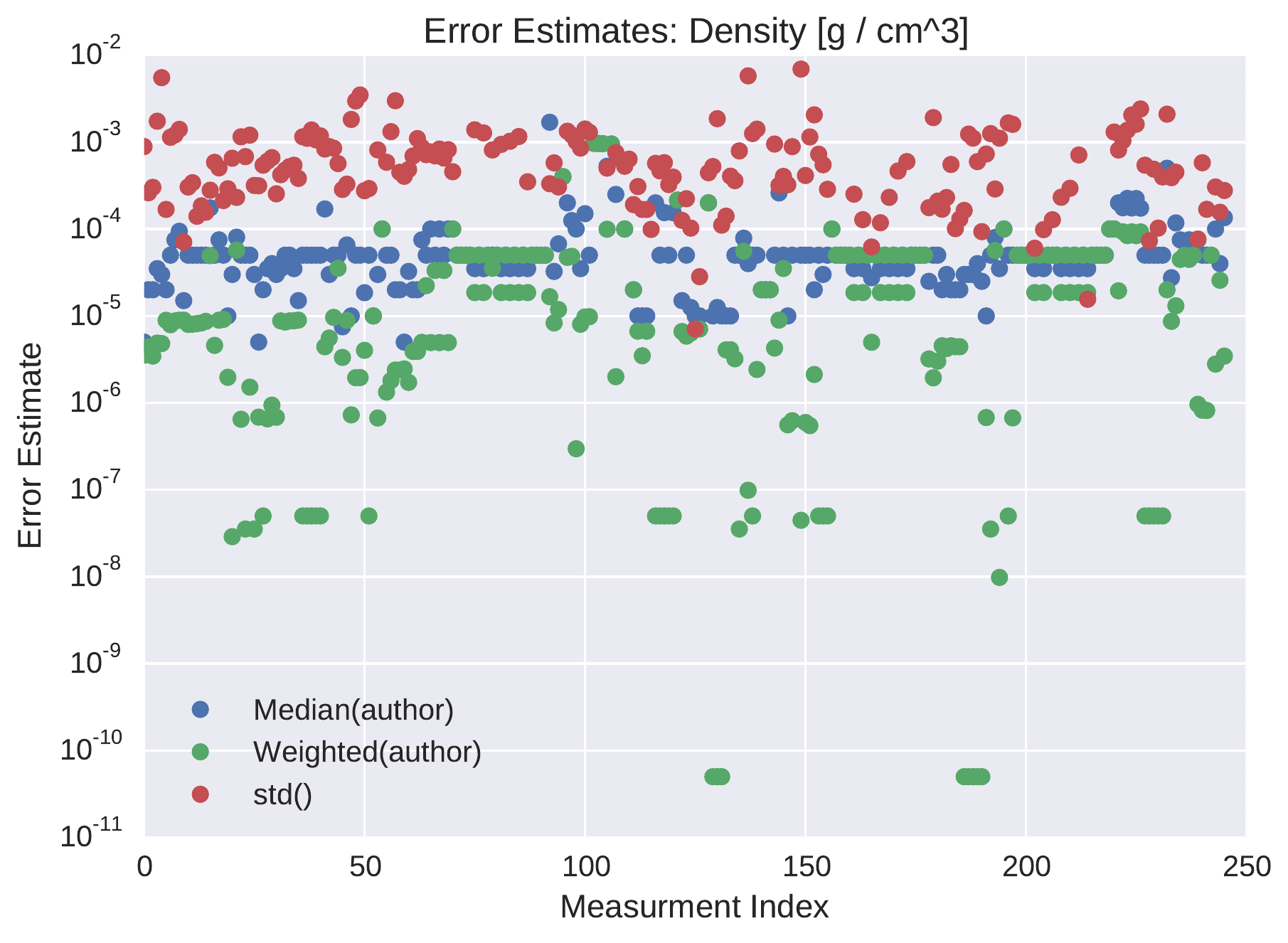}
}

\caption{{\bf Assessment of experimental error: Density}
To assess the experimental error in our ThermoML extract, we compared three different estimates of uncertainty.  
In the first approach (Weighted), we computed the standard deviation of the optimally weighted average of the measurements, using the uncertainties reported by authors ($\sigma_{Weighted} = [\sum_k \sigma_k^{-2}]^{-0.5}$).
This uncertainty estimator places the highest weights on measurements with small uncertainties and is therefore easily dominated by small outliers and uncertainty under-reporting.
In the second approach (Median), we estimated the median of the uncertainties reported by authors; this statistic should be robust to small and large outliers of author-reported uncertainties.
In the third approach (Std), we calculated at the standard deviation of independent measurements reported in the ThermoML extract, completely avoiding the author-reported uncertainties.
Plot (a) compares the three uncertai8nty estimates.
We see that author-reported uncertainties appear to be substantially smaller than the scatter between the observed measurements.
A simple psychological explanation might be that because density measurements are more routine, the authors simply report the repeatability stated by the manufacturer (e.g. 0.0001 g / $cm^{3}$ for a Mettler Toledo DM40~\cite{mettlertoledo}).  
However, this hardware limit is not achieved due to inconsistencies in sample preparation and experimental conditions; see Appendix in Ref.~\cite{chirico2013improvement}.  
Panel (b) shows the same information as (a) but as a function of the measurement index, rather than as a scatter plot---because not all measurements have author-supplied uncertainties, panel (c) contains slightly more data points than (a, b).  
}
\label{figure:ErrorAnalysisDensity}

\end{figure}


\clearpage

\begin{figure}

\subfigure[]{
\includegraphics[width=7.3cm]{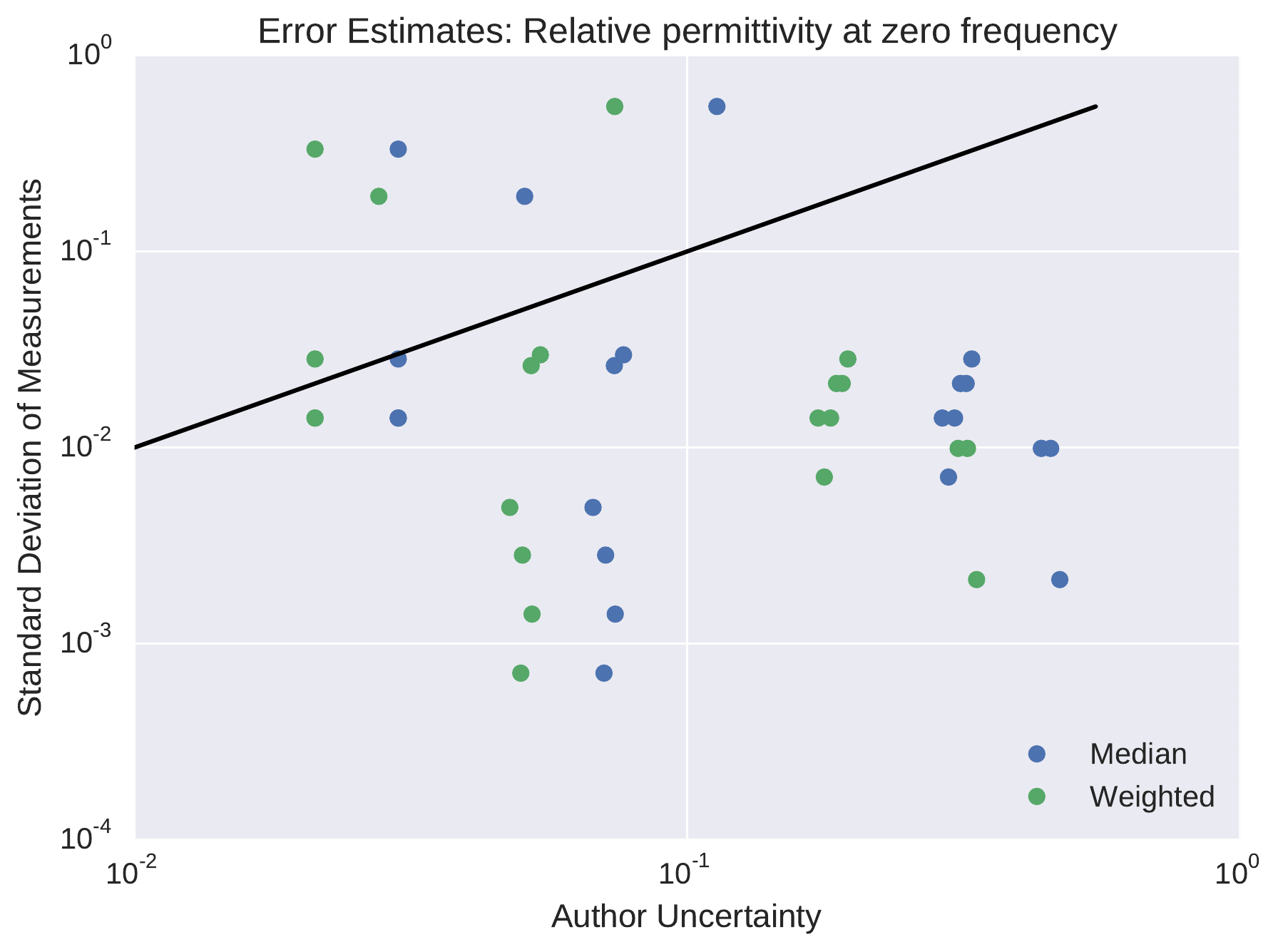}
}

\subfigure[]{
\includegraphics[width=7.3cm]{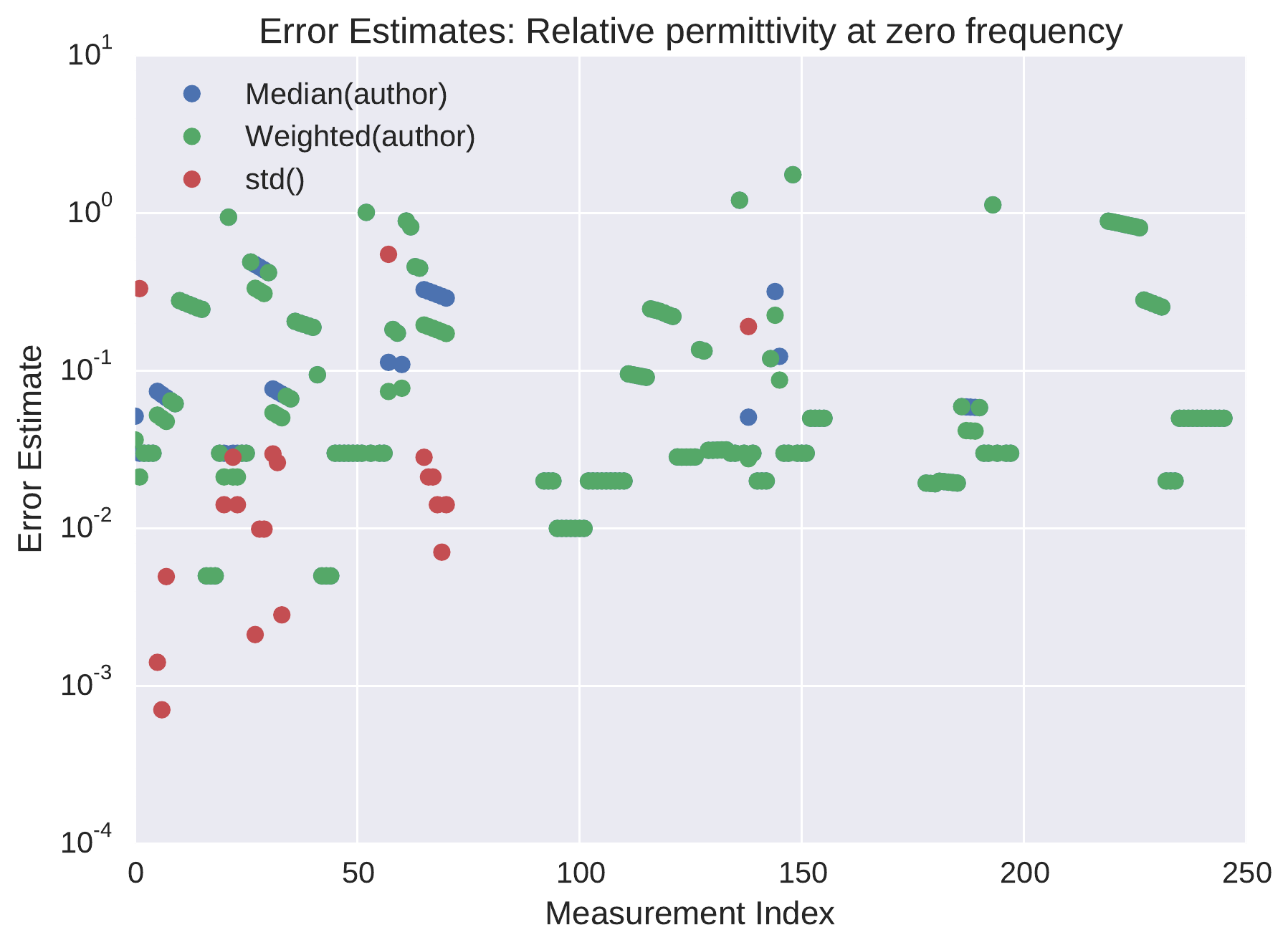}
}

\caption{{\bf Assessment of experimental error: Static Dielectric Constant}
To assess the experimental error in our ThermoML extract, we compared three different estimates of uncertainty.  
In the first approach (Weighted), we computed the standard deviation of the optimally weighted average of the measurements, using the uncertainties reported by authors ($\sigma_{Weighted} = [\sum_k \sigma_k^{-2}]^{-0.5}$).
This uncertainty estimator places the highest weights on measurements with small uncertainties and is therefore easily dominated by small outliers and uncertainty under-reporting.
In the second approach (Median), we estimated the median of the uncertainties reported by authors; this statistic should be robust to small and large outliers of author-reported uncertainties.
In the third approach (Std), we calculated at the standard deviation of independent measurements reported in the ThermoML extract, completely avoiding the author-reported uncertainties.
Plot (a) compares the three uncertainty estimates.
Unlike the case of densities, author-reported uncertainties appear to be somewhat larger than the scatter between the observed measurements.
Panel (b) shows the same information as (a) but as a function of the measurement index, rather than as a scatter plot---because not all measurements have author-supplied uncertainties, panel (c) contains slightly more data points than (a, b).  
}
\label{figure:ErrorAnalysisDielectric}

\end{figure}

\begin{figure*}[alldensity]

\includegraphics[width=\textwidth]{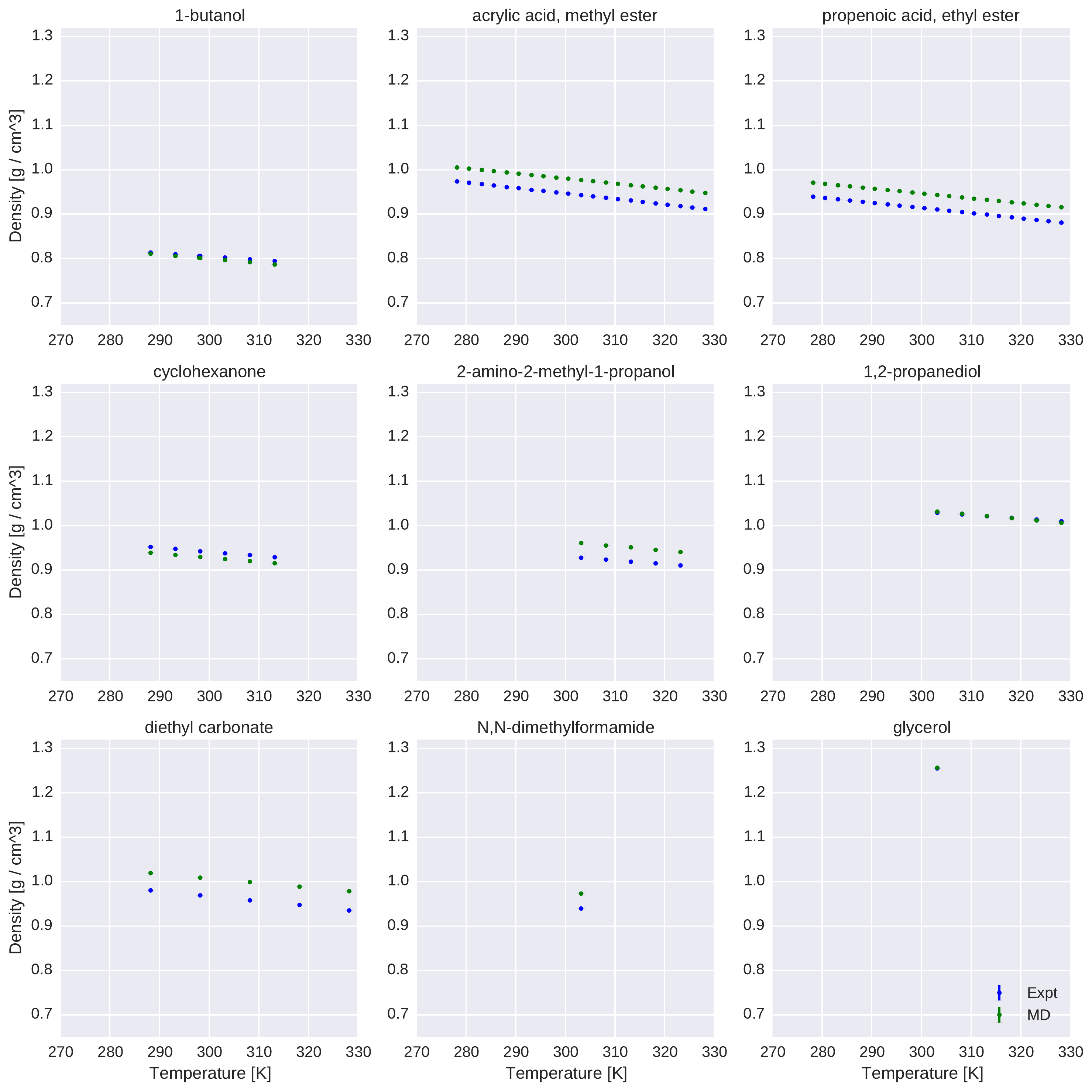}

\caption{{\bf Comparison of simulated and experimental densities for all compounds.} 
Measured (blue) and simulated (green) densities are shown in units of g / $cm^{3}$.
\label{figure:AllDensities}
}

\end{figure*}

\begin{figure*}[alldensity]

\ContinuedFloat

\includegraphics[width=\textwidth]{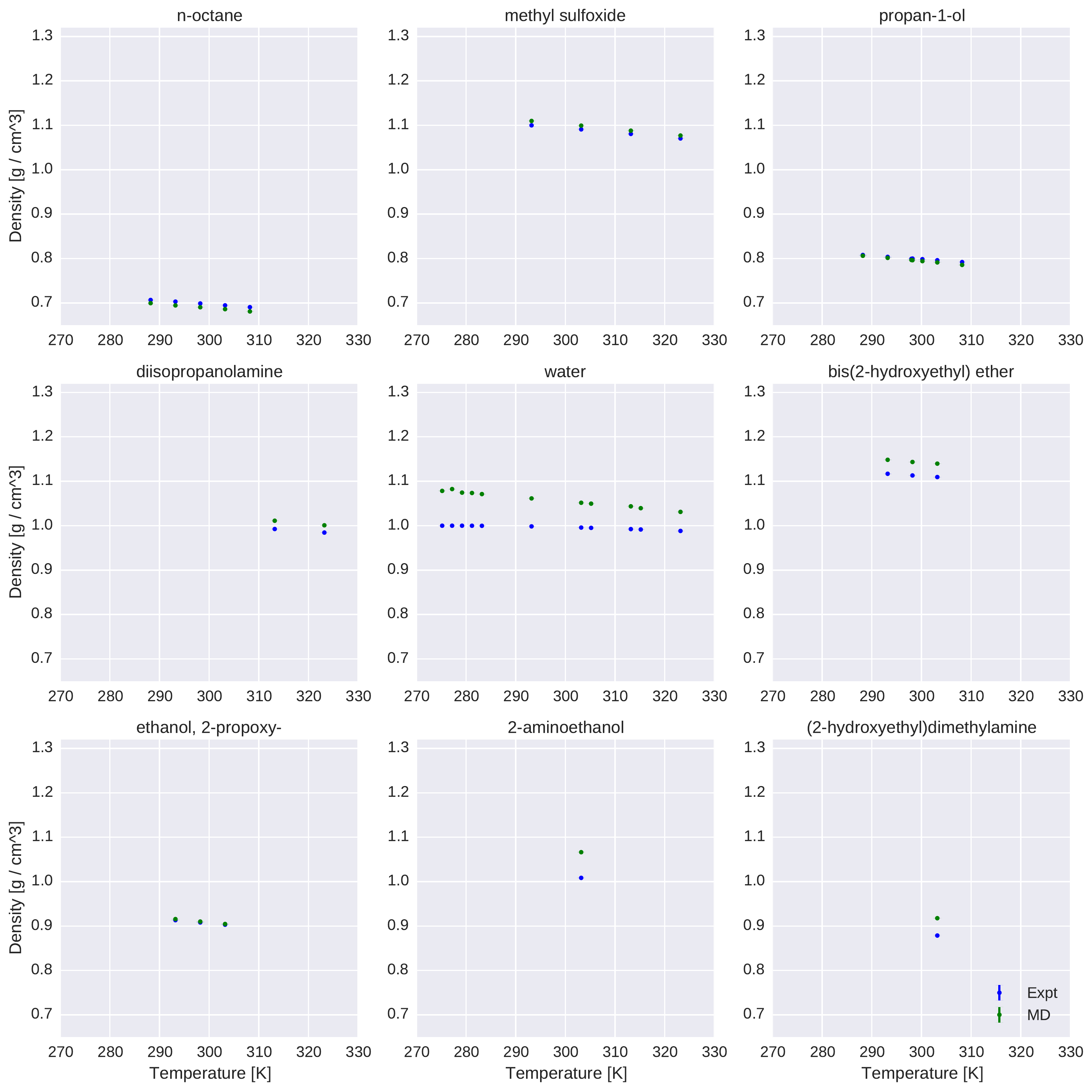}

\caption{{\bf Comparison of simulated and experimental densities for all compounds.} 
Measured (blue) and simulated (green) densities are shown in units of g / $cm^{3}$.
\label{figure:AllDensities}
}

\end{figure*}

\begin{figure*}[alldensity]

\ContinuedFloat

\includegraphics[width=\textwidth]{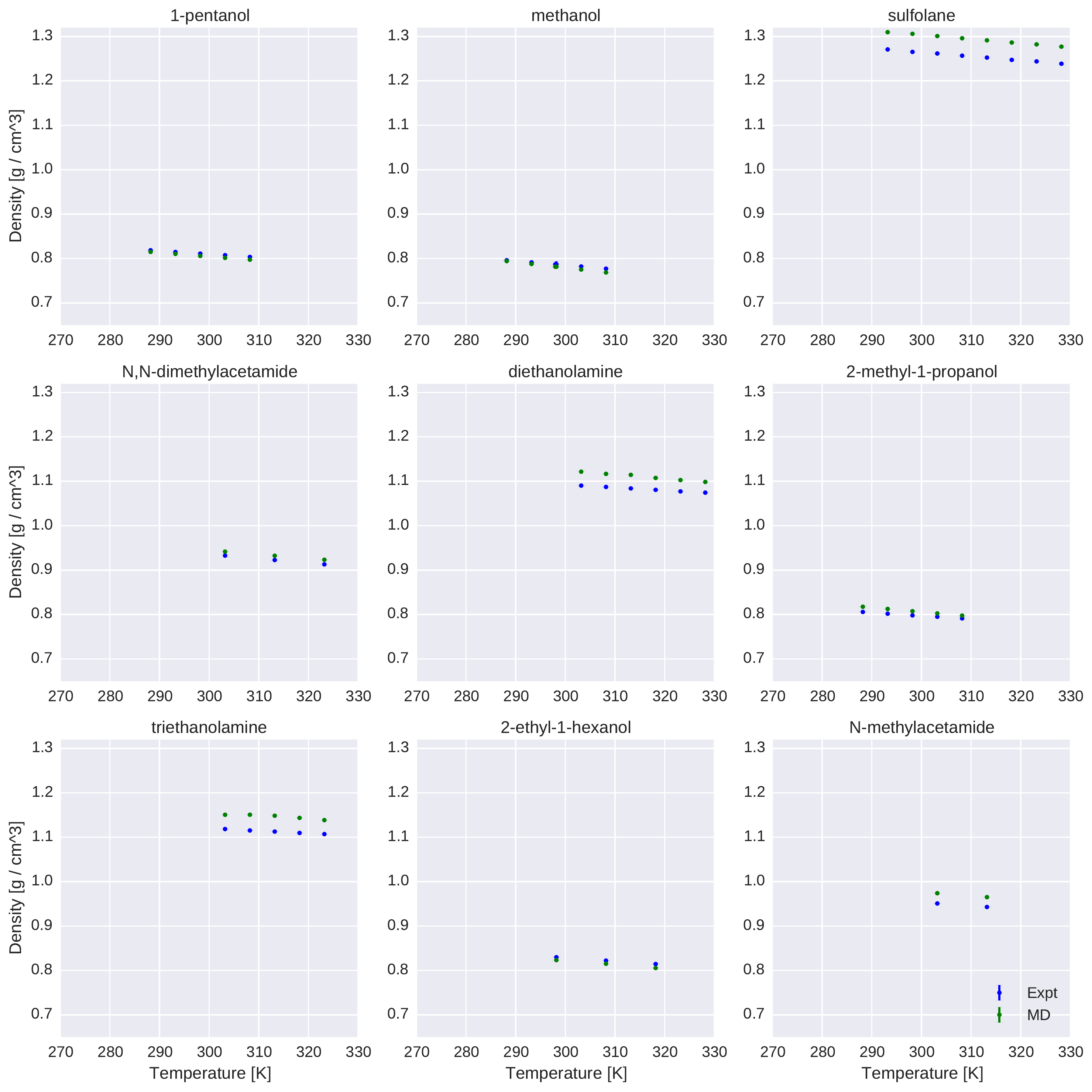}

\caption{{\bf Comparison of simulated and experimental densities for all compounds.} 
Measured (blue) and simulated (green) densities are shown in units of g / $cm^{3}$.
\label{figure:AllDensities}
}

\end{figure*}

\begin{figure*}[alldensity]

\ContinuedFloat

\includegraphics[width=\textwidth]{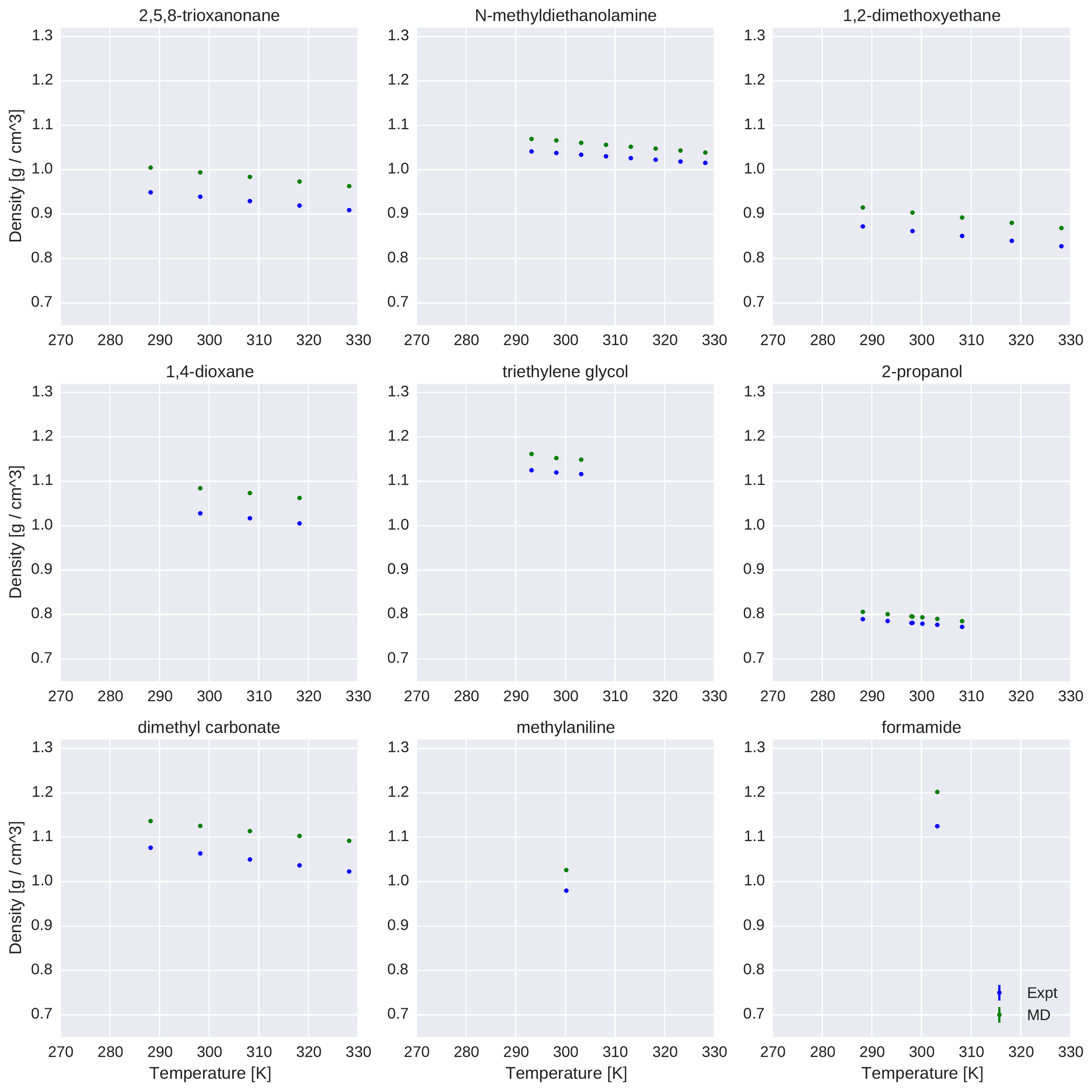}

\caption{{\bf Comparison of simulated and experimental densities for all compounds.} 
Measured (blue) and simulated (green) densities are shown in units of g / $cm^{3}$.
\label{figure:AllDensities}
}

\end{figure*}

\begin{figure*}[alldensity]

\ContinuedFloat

\includegraphics[width=\textwidth]{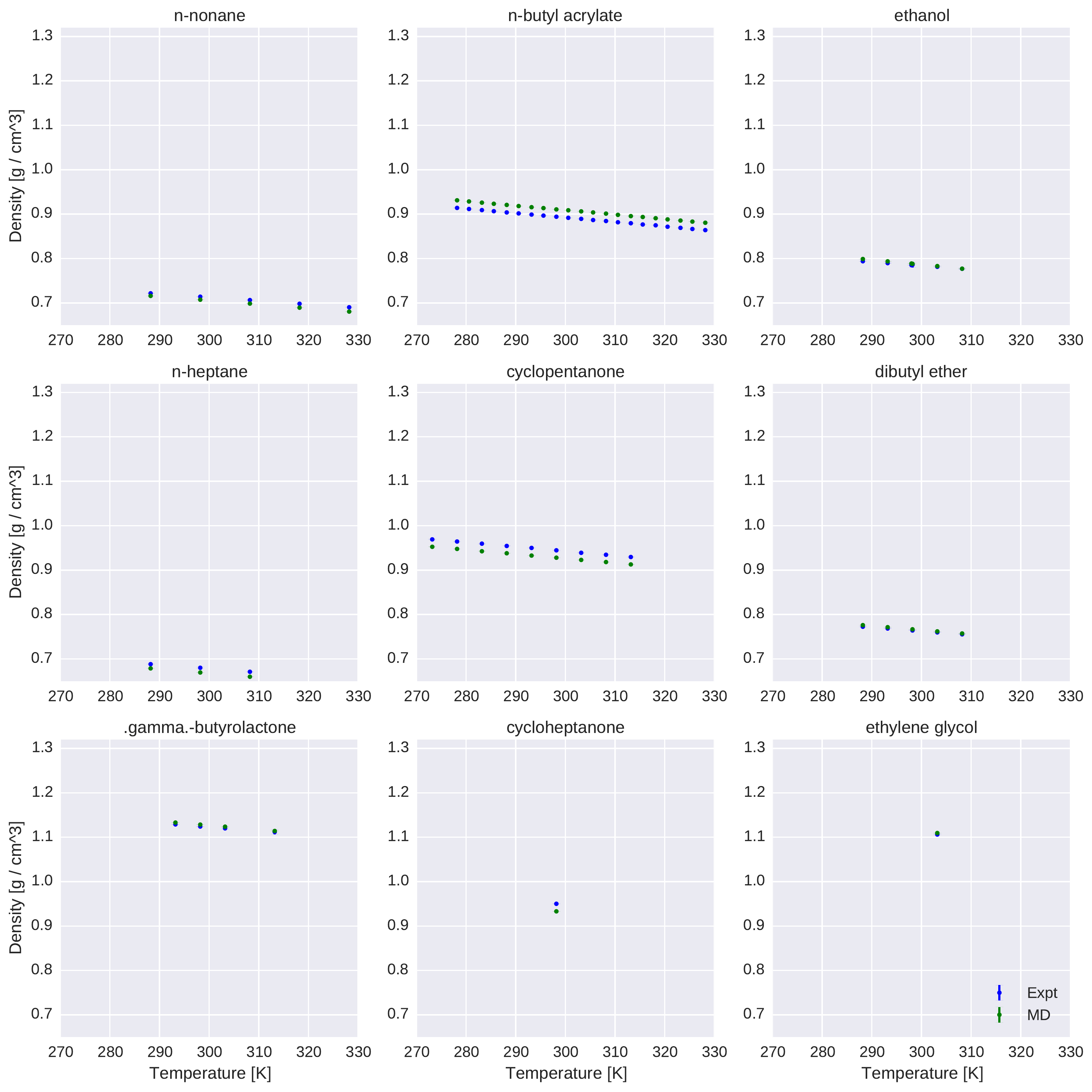}

\caption{{\bf Comparison of simulated and experimental densities for all compounds.} 
Measured (blue) and simulated (green) densities are shown in units of g / $cm^{3}$.
\label{figure:AllDensities}
}

\end{figure*}


\begin{figure*}[alldielectric]

\includegraphics[width=\textwidth]{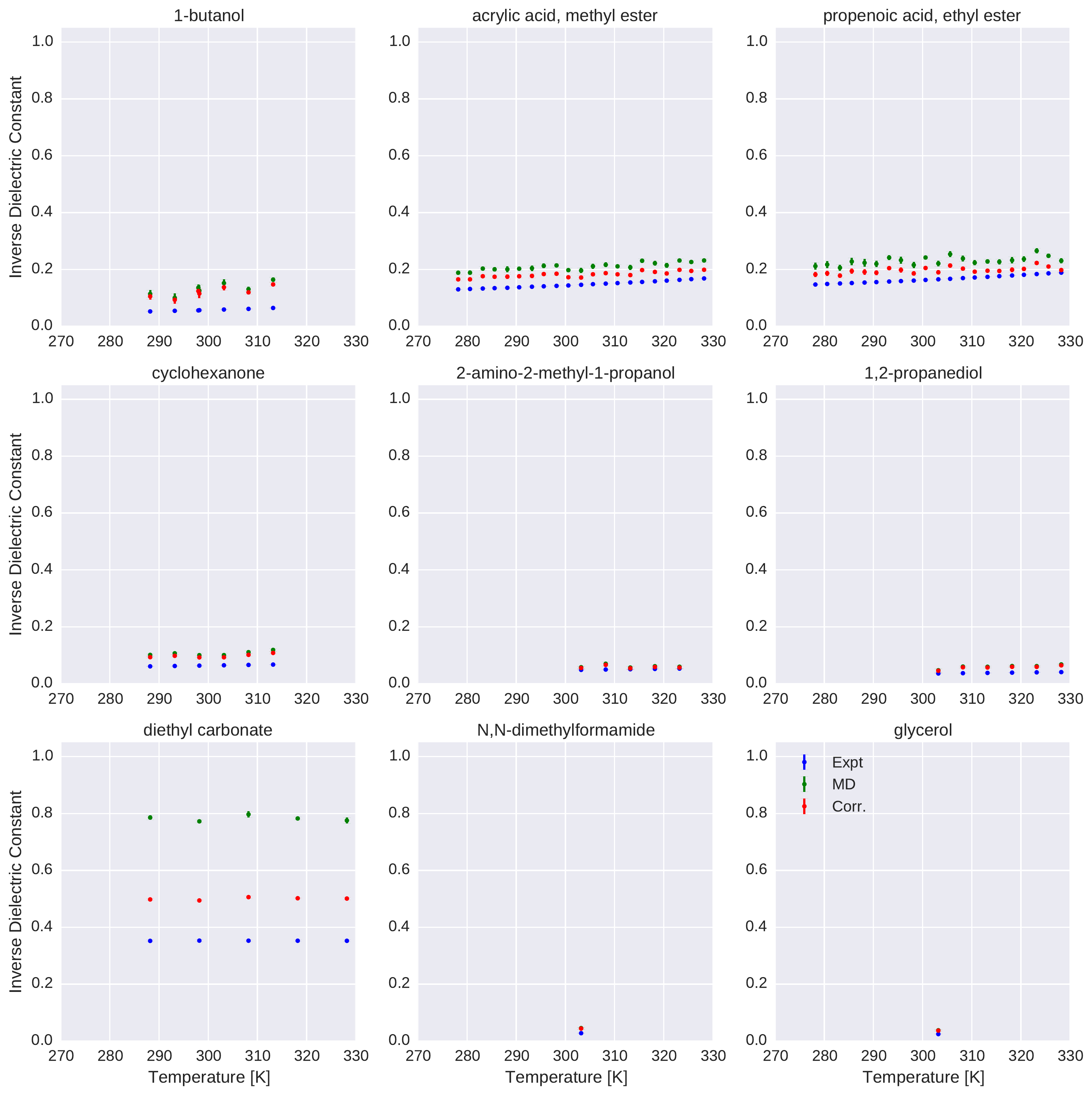}

\caption{{\bf Comparison of simulated and experimental static dielectric constants for all compounds.}
Measured (blue), simulated (green), and polarizability-corrected simulated (red) static dielectric constants are shown for all compounds.
}

\label{figure:AllDielectrics}

\end{figure*}

\begin{figure*}[alldielectric]

\ContinuedFloat

\includegraphics[width=\textwidth]{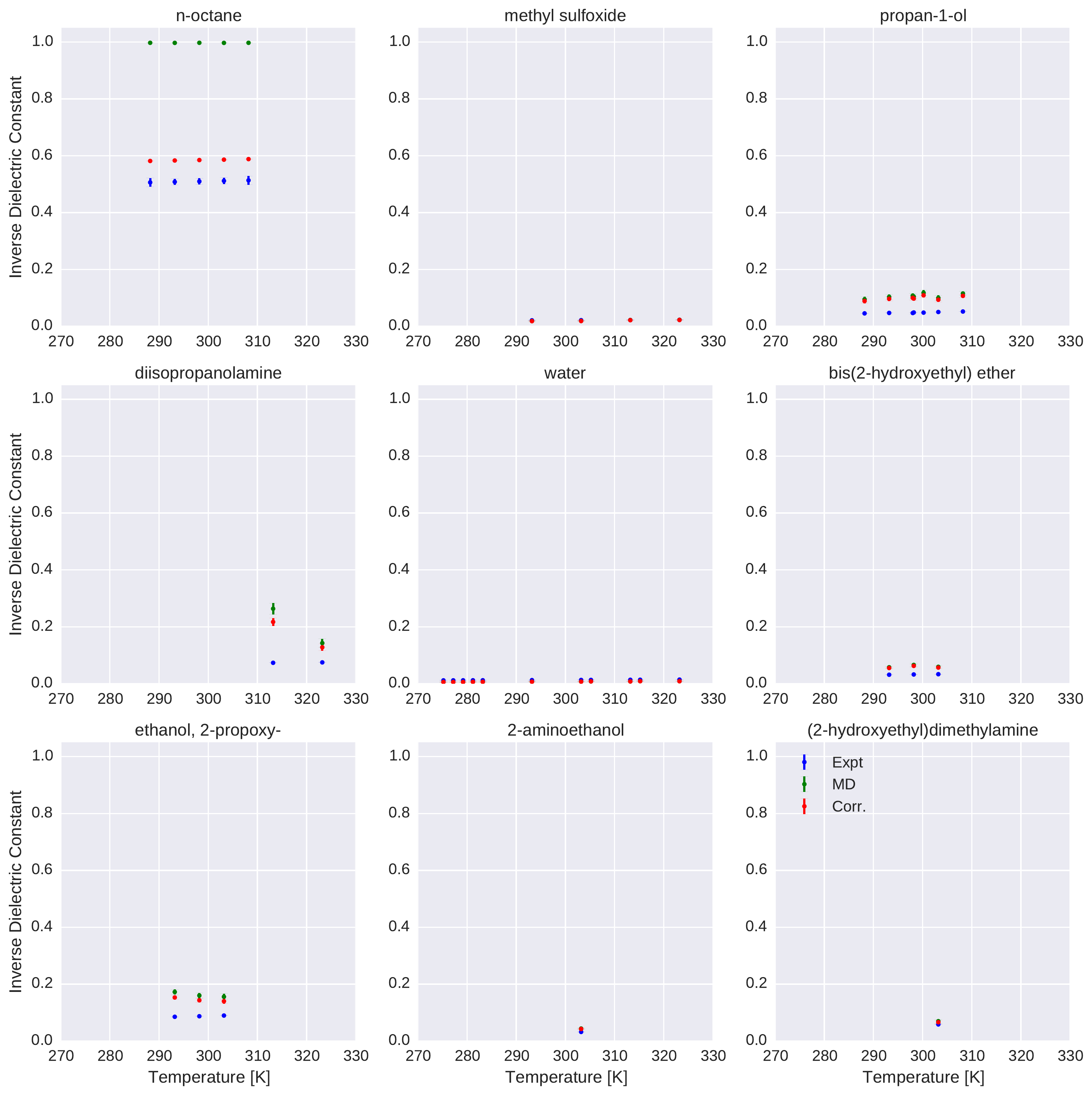}

\caption{{\bf Comparison of simulated and experimental static dielectric constants for all compounds.}
Measured (blue), simulated (green), and polarizability-corrected simulated (red) static dielectric constants are shown for all compounds.
}

\label{figure:AllDielectrics}

\end{figure*}

\begin{figure*}[alldielectric]

\ContinuedFloat

\includegraphics[width=\textwidth]{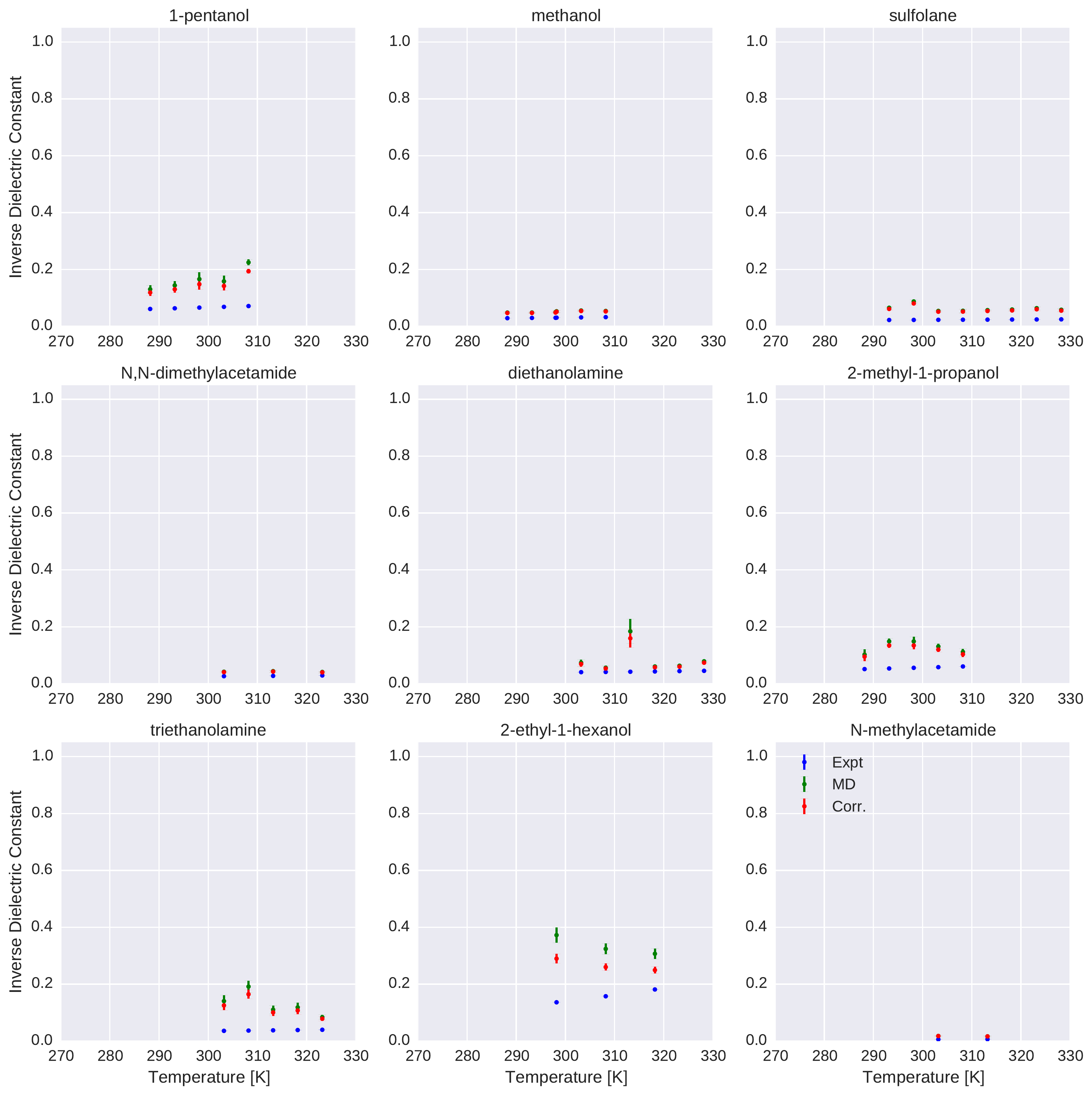}

\caption{{\bf Comparison of simulated and experimental static dielectric constants for all compounds.}
Measured (blue), simulated (green), and polarizability-corrected simulated (red) static dielectric constants are shown for all compounds.
}

\label{figure:AllDielectrics}

\end{figure*}

\begin{figure*}[alldielectric]

\ContinuedFloat

\includegraphics[width=\textwidth]{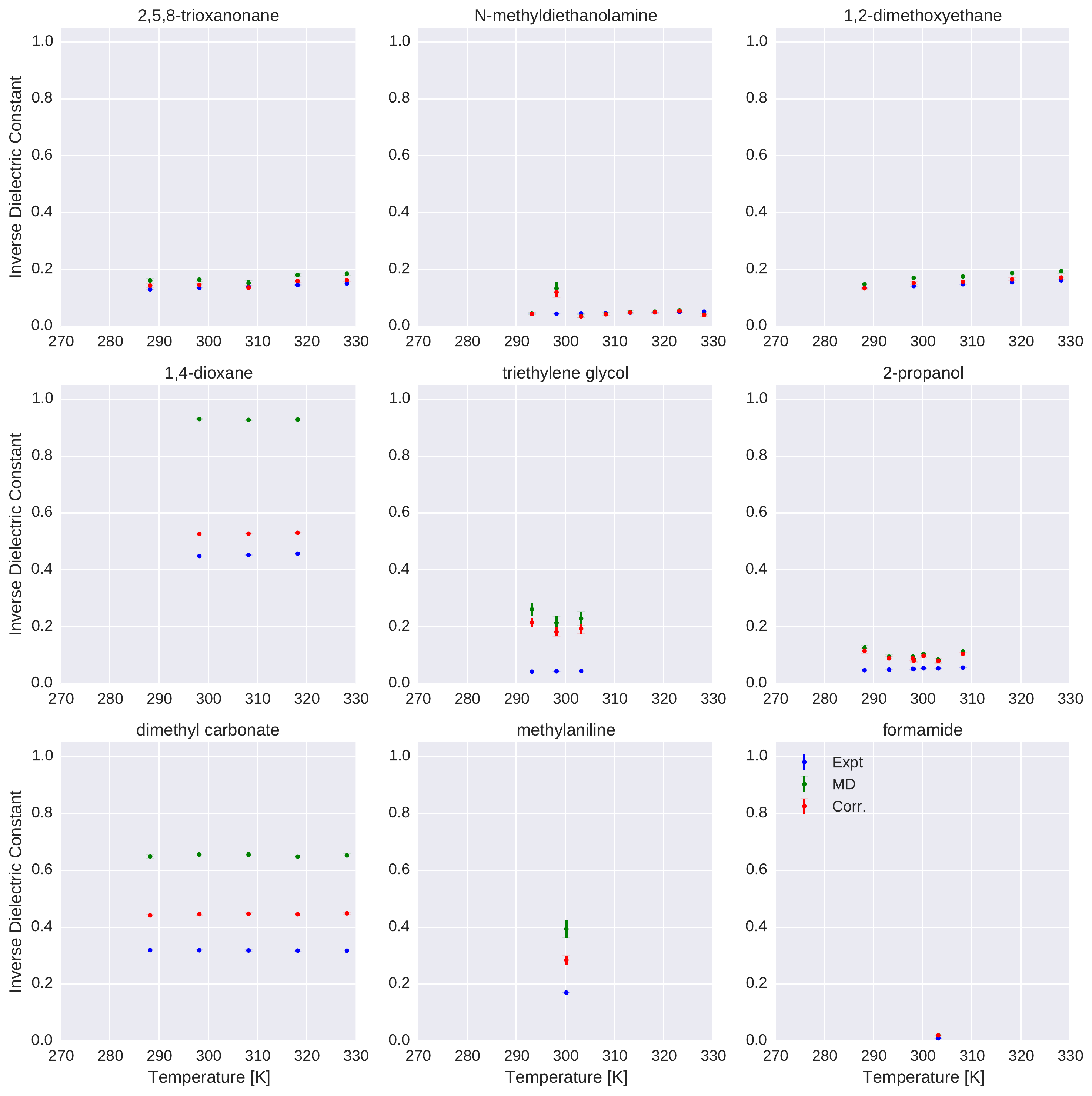}

\caption{{\bf Comparison of simulated and experimental static dielectric constants for all compounds.}
Measured (blue), simulated (green), and polarizability-corrected simulated (red) static dielectric constants are shown for all compounds.
}

\label{figure:AllDielectrics}

\end{figure*}

\begin{figure*}[alldielectric]

\ContinuedFloat

\includegraphics[width=\textwidth]{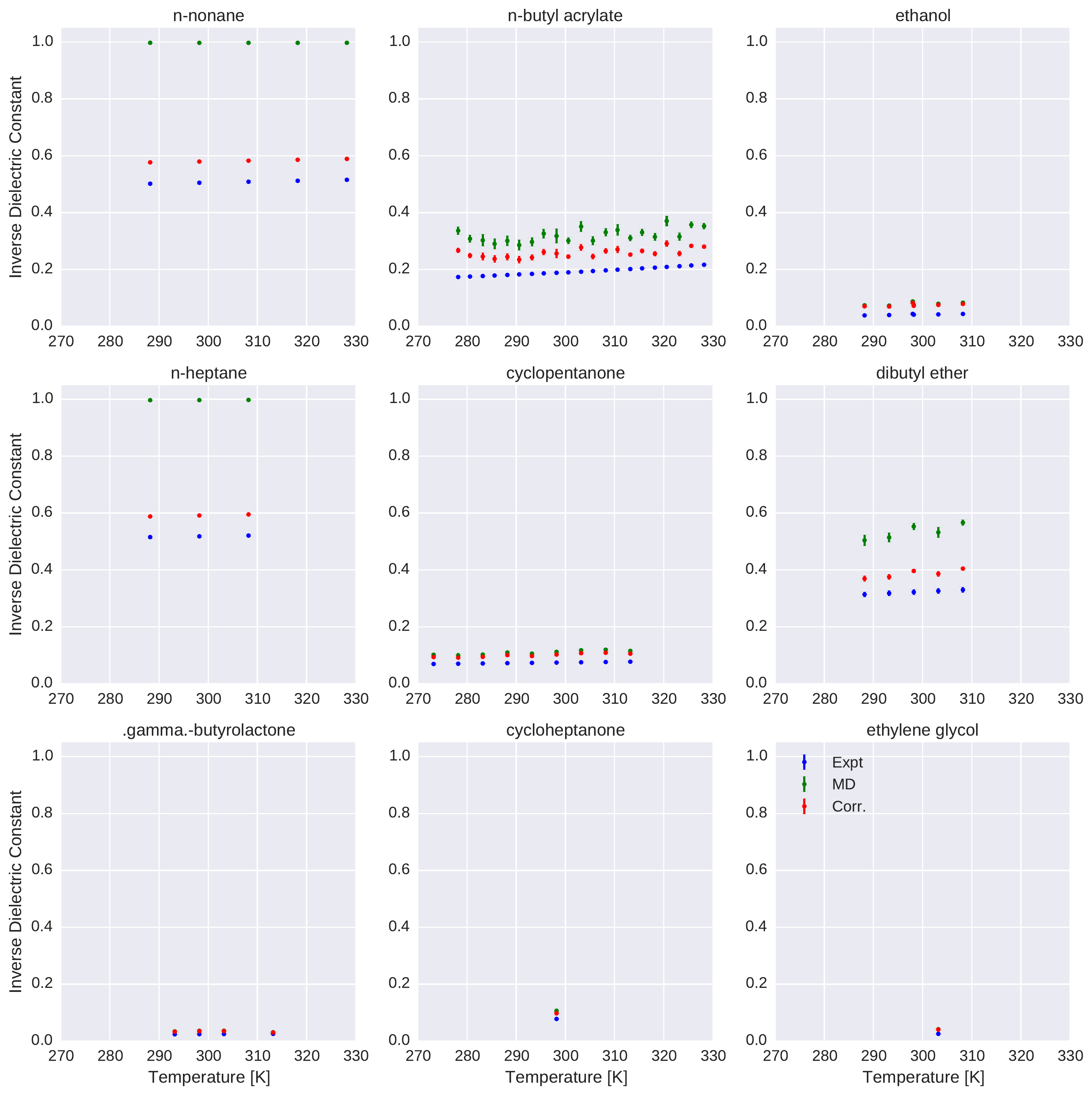}

\caption{{\bf Comparison of simulated and experimental static dielectric constants for all compounds.}
Measured (blue), simulated (green), and polarizability-corrected simulated (red) static dielectric constants are shown for all compounds.
}

\label{figure:AllDielectrics}

\end{figure*}

\clearpage

\subsection{Dependency Installation}
\label{section:commands}

The following shell commands can be used to install the necessary prerequisites via the {\tt conda} package manager for Python:

\begin{lstlisting}[language=bash]
$ conda config --add channels http://conda.binstar.org/omnia
$ conda install "openmoltools" "pymbar==2.1" "mdtraj==1.3" "openmm==6.3" packmol
%\end{lstlisting}

Note that this command installs the exact versions used in the present study, with the exception of openmoltools for which only a more recent package is available.  
However, for authors interested in extending the present work, we suggust using the most up-to-date versions available instead, which involves replace the equality symbols == with >=.

\clearpage

\bibliography{benchmark}

\end{document}